\documentclass[11pt,a4paper]{article}
\usepackage[utf8]{inputenc}
\usepackage{geometry}
\usepackage{graphicx}
\usepackage{array}
\usepackage{subfig}
\usepackage{slashed}
\usepackage{amsmath}
\usepackage{amsfonts}
\usepackage{amssymb}
\usepackage{tensor}
\usepackage{cite}
\usepackage{epstopdf}
\usepackage{comment}
\usepackage{hyperref}
\usepackage{cancel}
\usepackage{mathrsfs}

\usepackage{cleveref}
\crefname{section}{§}{§§}
\Crefname{section}{§}{§§}

\newcommand{\be}{\begin{equation}}
\newcommand{\ee}{\end{equation}}

\newcommand{\beq}{\begin{equation}}  \newcommand{\eeq}{\end{equation}}
\newcommand{\bal}{\begin{aligned}}   \newcommand{\eal}{\end{aligned}}
\newcommand{\bea}{\begin{eqnarray}}  \newcommand{\eea}{\end{eqnarray}}
\def\d{\mathrm{d}}
\def\t{t}

\begin{document}

\begin{titlepage}

\begin{flushright}
%\vskipcm
%{\small  CERN--TH--2018--144},
{\small  MPP--2019--198},
{\small LMU--ASC 32/19},
{\small IPhT-T19/133},
{\small CPHT-RR055.092019}
\end{flushright}

\vspace{0.6cm}

\begin{center}

{\LARGE{\textsc{Swampland, Gradient Flow and Infinite Distance}}}

%\vskip0.4cm

%{\LARGE{\textsc{its Holographic Description from Superstrings}}}

\vskip0.5cm

%{\LARGE{\textsc{Weyl Supergravity}}}

%\vskip0.4cm

% {\LARGE{\textsc{S-fold and Double Copy Construction}}}

\vspace{1cm}
{\large \bf Alex Kehagias$^{a}$,\, Dieter L\"ust$^{b,c}$,\, Severin L\"ust$^{d,e}$}

\vspace{1cm}

{\it

\vskip-0.3cm
\centerline{ $^{\textrm{a}}$ Physics Division, National Technical University of Athens}
\centerline{ 15780 Zografou Campus, Athens, Greece}
\medskip
\centerline{ $^{\textrm{b}}$ Arnold--Sommerfeld--Center for Theoretical Physics,}
\centerline{Ludwig--Maximilians--Universit\"at, 80333 M\"unchen, Germany}
\medskip
\centerline{$^{\textrm{c}}$ Max--Planck--Institut f\"ur Physik,
Werner--Heisenberg--Institut,}
\centerline{ 80805 M\"unchen, Germany}
\medskip
\centerline{$^{\textrm{d}}$ Institut de Physique Th\'eorique, 
Universit\'e Paris Saclay, CEA, CNRS}
\centerline{Orme des Merisiers}
\centerline{91191 Gif-sur-Yvette Cedex, France}
\medskip
\centerline{$^{\textrm{e}}$ CPHT, CNRS, Ecole Polytechnique, IP Paris,} 
\centerline{91128 Palaiseau Cedex, France}

}

\end{center}

\vskip0.5cm
\abstract{{
In the first part of this paper we will work out a close and so far not yet noticed correspondence between the swampland approach in quantum gravity 
and geometric flow equations in general relativity, most notably the
Ricci flow.
We conjecture that following the gradient flow towards a fixed point, which is at infinite distance in the space of background metrics,
is accompanied by an infinite tower of states in quantum gravity. In case of the Ricci flow, this conjecture is in accordance with the generalized distance and AdS distance conjectures, 
 which were recently discussed in the literature, but it should also hold
 for more general background spaces.  We argue that the entropy functionals of  gradient flows
provide a useful definition of the generalized distance in the space of background fields. 
In particular we give evidence that for the Ricci flow the distance $\Delta$ can be defined in terms of
the mean scalar curvature of the manifold, $\Delta\sim\log \bar R$. For a more general gradient flow, the distance functional also depends on the string coupling constant.

In the second part of the paper we will apply the generalized distance conjecture to gravity theories with 
higher curvature interactions, like higher derivative $R^2$ and $W^2$ terms. We will show that going to the weak coupling limit of the higher derivative terms corresponds to the infinite distance limit
in metric space
and hence this
limit must be accompanied by an infinite tower of light states. For the case of the $R^2$ or  $W^2$ couplings, this limit corresponds to the limit of a small cosmological constant or, respectively,
 to a light additional spin-two field in gravity. 
 In general we see that the limit of small  higher curvature couplings belongs to the swampland in quantum gravity,
 just like the limit of a small $U(1)$  gauge coupling belongs to the swampland as well.
}}

\end{titlepage}

\newpage

\section{Introduction}

Recently the so-called swampland approach \cite{Vafa:2005ui,Ooguri:2006in,Palti:2019pca}
addressed the important question of what kind of effective field theories, albeit being apparently fully consistent from the quantum field theory point of view,
cannot be consistently embedded into quantum gravity. During the course of this discussion, it turned out that  directions in the field space, which are associated to large field values and to weak coupling in
quantum field theory, are rejected in quantum gravity in the sense that they are accompanied by an infinite  tower of almost massless particles. Hence the field theory cutoff scale  has to be as low as the mass scale of the tower itself and eventually becomes zero for infinite field values, where the effective field description is  invalidated at all.
One of the best understood examples of the swampland idea is the weak gravity conjecture  \cite{ArkaniHamed:2006dz},
stating that a $U(1)$ gauge theory with very small gauge coupling constant is accompanied by a tower of very light states.

The swampland idea together with the infinite
distance conjecture was tested  in string theory in many interesting instances 
\cite{Klaewer:2016kiy,Heidenreich:2017sim,Andriolo:2018lvp,Grimm:2018ohb,Heidenreich:2018kpg,Blumenhagen:2018nts,Blumenhagen:2018hsh,Lee:2018urn,Lee:2018spm,Grimm:2018cpv,Corvilain:2018lgw,Lee:2019tst,Joshi:2019nzi,Marchesano:2019ifh,Font:2019cxq,Lee:2019xtm,Erkinger:2019umg}.
In many cases so far one asks  the question, if particular effective matter 
quantum field theories like effective gauge theories  can be consistently coupled to quantum gravity or not. 
In this paper we will consider the problem, which kind of effective gravity theories belong to the swampland and which ones, on the other hand, constitute fully consistent quantum gravity theories.
Previous attempts in this direction are the (refined) de Sitter conjecture \cite{Obied:2018sgi,Ooguri:2018wrx}, bounds on slow roll inflation and the de Sitter swampland \cite{Garg:2018reu}
and the quantum exclusion of de Sitter space \cite{Dvali:2013eja,Dvali:2014gua,Dvali:2017eba,Dvali:2018fqu,Dvali:2018jhn},
which conjecture that quantum gravity with a positive cosmological constant is inconsistent.
For anti-de Sitter space,
 the AdS distance conjecture \cite{Lust:2019zwm}
states
that the limit of vanishing AdS cosmological constant $\Lambda\rightarrow 0$ is at infinite distance in field space and hence accompanied by an infinite tower of massless states. Another example \cite{Klaewer:2018yxi}, 
where the
swampland idea was utilized within gravity, are massive spin-two theories, leading to the conjecture that the limit of very light massive spin-two particles belongs to the swampland in the sense that it
is again vetoed by an infinite tower of light states.

The AdS distance conjecture is based on a specific generalization \cite{Lust:2019zwm} of the original distance conjecture, applying it not only to internal moduli fields but also to the d-dimensional space-time metric itself.
For the case of Weyl rescaling of the space-time metric the generalized distance conjecture then immediately leads to the AdS distance conjecture.

Looking at the generalized distance conjecture from a more general, geometric point of view, we will show that the definition of the distance within the space of background metrics is very closely related to 
mathematical flow equations in general relativity, where one follows the flow of a family of metrics with respect to a certain path in field space. The most famous example is the Ricci flow, introduced by Hamilton  \cite{hamilton},
and subsequently investigated, among others, by Perelman \cite{Perelman:2006un}. In the first part of this paper we will work out the close correspondence between the generalized distance conjecture and the
Ricci flow. We will  first show that for  curved Einstein spaces the fixed points of the Ricci flow with vanishing curvature are indeed at infinite distance in the field space with respect to a canonically defined distance, which
was already used in \cite{Lust:2019zwm}. Even more concretely, the Ricci flow in the cosmological constant $\Lambda$ leads to a fixed point with zero cosmological constant, which is infinitely far away from any finite value
of  $\Lambda$. Using this observation we conjecture that  following the Ricci flow towards a fixed point, which is at infinite distance in the background space,
is accompanied by an infinite tower of states in quantum gravity.
 We are testing  this conjecture also by several other examples. 
 We also discuss the role of the Ricci flow in a proper string theory embedding and also for large distance "flows" in the moduli space of Calabi-Yau manifolds.
 
 A second important point,
we conjecture  and give evidence for, is that the entropy functionals of the gradient flow equations provide a sensible definition for the distance 
and hence for corresponding masses of the tower
of states along the flow.
For the case of the Ricci flow, the relevant distance can be defined in terms of the  scalar curvature of the background metric: $\Delta\simeq\log R$.
Considering the combined dilaton-metric flow, the corresponding grading flow is derived from the entropy functional ${\cal F}$. As we will see ${\cal F}$ provides a good definition for the distance in the combined field space 
of the background metric and of the string coupling constant $g_s$.
This argument will be again augmented by an explicit example related to two-dimensional $\lambda$-deformed backgrounds \cite{sf1,sf2}
and the two-dimensional black hole solution 
\cite{Witten:1991yr}.
Finally, the entropy functional ${\cal W}$
of Perelman provides a proper definition of the distance for the Perelman flow.

The use of the geometric flow equations  provides a new and more geometric understanding of the infinite distance conjecture within the swampland approach.
But this connection also puts the gradient flows in mathematics into a quantum gravity perspective, in the sense that it shows
that approaching infinite distance fixed points of the gradient flow the associated gravity theories always need a quantum gravity completion in the form of extra light quantum states.

In the second part of the paper we consider the generalized distance conjecture, applied for Weyl rescalings of the external space-time metric in the context of
effective d-dimensional gravity theories that include higher curvature interactions. Concretely, we will argue that the associated higher curvature coupling constants exhibit an infinite distance behaviour
with respect to Weyl rescalings of the d-dimensional metric. This means that whenever one of these couplings goes to zero (or infinity) an infinite tower of states becomes massless.
This behaviour results entirely from an infinite distance with respect to the d-dimensional external metric and applies with
no reference to any internal space. However the associated tower of massless states often originates from some extra internal compact space, like KK modes of the internal space.\footnote{But it could also be that
the infinite tower has no interpretation in terms of  extra dimensions.}

We will also argue  that already the limit of vanishing gauge coupling constants of an effective $U(1)$ gauge theory
is at infinite distance with respect to Weyl rescalings of the external metric. 
Then we will discuss in more detail first
quadratic gravity theory with $R^2$ action. The weak coupling limit  with respect to the  $R^2$ coupling constant will be shown to be at infinite distance; this behaviour is shown to be equivalent to the  AdS distance conjecture
of the cosmological constant. 
Similarly, for the quadratic Weyl-square theory with $W^2$ action,
we will see that the weak coupling limit   is at infinite distance
with respect to Weyl rescalings. Now this behaviour is equivalent to the spin-two swampland conjecture \cite{Klaewer:2018yxi}, stating that the limit of vanishing spin-two mass
is at infinite distance, being   always associated with an infinite tower of light states.
As we will discuss in addition, considering  the $W^2$-theory on an (A)dS background, one obtains a theory with a partially massless spin-two field. 
As already mentioned in \cite{Lust:2019lmq},
we will also
show that the Higuchi limit \cite{Higuchi:1986py}
is at infinite distance with respect to Weyl rescaling of the (A)dS metric.

\section{Einstein gravity, the generalized distance conjecture and Ricci flow}

\subsection*{The generalized distance conjecture}

Let us recall the generalized distance conjecture \cite{Lust:2019zwm}: in the following
we consider  d-dimensional manifolds $M_d$ with metric variations $g_{\mu\nu}(\tau)$, where $\tau$ parametrizes a family of metrics of $M_d$.
This means that  we do not apply the distance conjecture
to the variations of some matter fields, like the scalar moduli of an internal Calabi-Yau manifold, but to the variations of the space-time metric itself.
We consider  a geodesic path $\gamma$ in field space, which is parametrised by $\tau$ ranging from an initial $\tau_i$ to a final $\tau_f$,   so that the proper metric distance $\Delta$ along the path is
 \cite{DeWitt,Michor}
\be
\Delta_g = c \int_{\tau_i}^{\tau_f}  \left( \frac{1}{V_M} \int_M \sqrt{g} g^{MN} g^{OP} \frac{\partial g_{MO}}{\partial \tau} \frac{\partial g_{NP}}{\partial \tau}   \right)^{\frac12}d\tau \;,
\label{dismetr}
\ee
where $c \sim {\cal O}(1)$ is a constant depending on the dimension of $M$.\footnote{In     this distance formula was applied for transverse traceless metric variations. However it is also valid for 
general variations of the metric.}
Then the  generalized distance conjecture states that there must be an infinite tower of states with mass scale $m\left(\tau\right)$ such that 
\be
m\left(\tau_f\right) \sim m\left(\tau_i\right) e^{-\alpha | {\Delta_g }|} \,,
\label{dsc}
\ee 
where $\alpha \sim {\cal O}(1)$.
  Generically the masses $m\left(\tau_i\right)$ at the initial  value $\tau_i$ are of the order of the Planck mass, such that one obtains 
\be
m \sim M_p e^{-\alpha |\Delta_g| } \,.
\label{gendisc}
\ee
The mass scale $m$ is setting the natural cut-off space, above which the effective field theory description breaks down.
 Hence for large distance variations with $|\Delta_g|\rightarrow\infty$ in the space of metrics, we get in quantum gravity a massless tower of states, which invalidates the effective field theory description.
 In this sense the large distance limit is said to lead to effective field theories that belong to the swampland.

 Besides the general notion of distance  in the space of metrics, the so-called Weyl distance was especially considered in \cite{Lust:2019zwm}. 
It refers  to  distances in metric space associated to variations of the external metric that are Weyl rescalings.
Specifically, under an external Weyl variation of the external space with an exponential Weyl factor of the form
\be
\tilde{g}_{\mu\nu} = e^{2\tau} g_{\mu\nu}\;.
\label{weytra1}
\ee
Applying the distance formula (\ref{dismetr}) to the Weyl transformed metric, or equivalently determining 
the canonically normalised field for measuring the distance, one gets that the metric distance with respect to Weyl rescalings is  just given by the field $\tau$:
 \begin{equation}\Delta_g\simeq\tau_f-\tau_i\simeq\tau\, .
\end{equation}
According to the distance conjecture there is a corresponding  tower of states with masses that scale  as 
\be
m \sim e^{-\alpha \tau }\;.
\label{gendisca1}
\ee
These tower of states become light in the limit      $\Delta_g\simeq\tau\rightarrow\infty$. Note that there is also the opposite large distance limit where $\Delta_g\simeq\tau\rightarrow-\infty$.
In this second large distance limit it can happen that  a different, dual  tower of states becomes light, which scales as $
m' \sim e^{\alpha \tau }$. Whether there is only one or two tower of light states depends on the circumstances of the considered coupling (see also the discussion in section 
III).

\vskip0.3cm
As in \cite{Lust:2019zwm}, we can now consider a family of (A)dS vacua parametrised by a variation of the cosmological constant $\Lambda(\tau)$ from $\Lambda_i$ to $\Lambda_f$. 
Specifically, under Weyl rescaling (\ref{weytra1}) the AdS cosmological constant is transforming in the following way:
\be
\Lambda = -\frac12 \left(d-1\right)\left(d-2\right) M_p^2e^{-2\tau} \;.
\ee
It follows that the metric distance between an initial value $\Lambda_i$ and a final value $\Lambda_f$ is given as
\begin{equation}
\Delta_g \simeq \tau_f-\tau_i \simeq \log\Biggl({\Lambda_i\over\Lambda_f}\Biggr)\, .\label{lambdadistance}
\end{equation}
We see that the limit $\Lambda_f\rightarrow 0$ is at infinite distance with respect to Weyl rescalings of the (A)dS metric, since it corresponds to the limit $\tau\rightarrow\infty$.
Actually,
this behaviour can then be combined with the generalised distance conjecture (\ref{gendisca1}) to state the  
\vskip0.2cm

\noindent
{\bf AdS Distance Conjecture  :}
{\it 
In quantum gravity on a $d$-dimensional AdS space with cosmological constant $\Lambda$, there exists an infinite tower of states with mass scale $m$ which, as $\Lambda \rightarrow 0$, behaves (in Planck units) as
\be
\label{dsdc}
m \sim |\Lambda|^{\alpha}  \;,
\ee 
where $\alpha$ is a positive order-one number.}

\vskip0.2cm
Furthermore one expects that for AdS spaces that  $\alpha\geq1/2$.
The same kind of distance conjecture can be also formulated for positive cosmological constant, i.e. in de Sitter space, if it exists in string theory or in quantum gravity.
In fact, it was argued in \cite{Lust:2019zwm} that for dS space the same relation between $\Lambda$ and $m$ holds, but with $0\leq\alpha\leq1/2$.\footnote{It might be tempting to
speculate
 that there is a kind of duality between AdS and dS space, namely an exchange of the towers $m_{AdS} \leftrightarrow m_{dS}$ when exchanging
$(- \Lambda_{AdS})^{1/2+ \alpha'} \leftrightarrow ( \Lambda_{dS})^{1/2- \alpha'}$.
%Moreover this duality could imply  that the AdS tower becomes very heavy for $\Lambda_{dS} \rightarrow 0$ and 
%and the dS tower becomes very heavy for $\Lambda_{AdS} \rightarrow 0$.
}

Formally one could also
consider a "dual" infinite distance limit with $\tau\rightarrow-\infty$. However this limit leads to a very large cosmological constant, $\Lambda\rightarrow \infty$ in Planck units, and the associated curvature of space time
is also super-Planckian. It is
not clear, if a tower of light states appears in this large curvature limit. Therefore we discard this non-geometric regime for the moment, keeping $\tau\geq0$, but we will come back to 
the limit $\tau\rightarrow-\infty$  in string theory, in M-theory and also in section III, when we discuss various coupling constants.

\subsection{Ricci flow and infinite distance}

Let us now consider  Einstein gravity on d-dimensional spaces $M_d$
and a specified  family of background metrics $g_{\mu\nu}(t)$, where now $t$ is the  parameter, along we want to consider the Ricci flow.
The Ricci flow equation, introduced by Hamilton  \cite{hamilton},  is a particular partial differential equation, which
relates the change of the metric to the Ricci tensor $R_{\mu\nu}$ of the associated manifold:
\begin{equation}
{\partial \over\partial t}g_{\mu\nu}(t)=-2R_{\mu\nu}(t). \label{rf}
\end{equation}
Note that $t$ here has mass dimensions $-2$ and it is different from the parameter $\tau$  parametrizing the Weyl rescaling in (\ref{weytra1}). 
Notice also that $t$ should not be confused with the physical time coordinate. 
Ricci flow implies that for positive Ricci curvature the manifold is contracting and for regions of negative curvature the manifold is extending. 
Of particular interest are the fixed points $\hat g_{\mu\nu}(\hat t)$ of the Ricci flow, namely those points, where the Ricci  flow is ending. These fixed points satisfy 
\begin{equation}
{\partial \over\partial t}g_{\mu\nu}(t)|_{t=\hat t}=0\, . 
\label{rfa}
\end{equation}
Hence the Ricci tensor is zero at the fixed point of the Ricci flow:
\begin{equation}
\hat R_{\mu\nu}(\hat t)=0\, . \label{flat}
\end{equation}
From this equation we see that the fixed point of the Ricci flow is flat space-time with vanishing curvature and vanishing cosmological constant.
On the other hand, as discussed in \cite{Lust:2019zwm}, the  metric with vanishing cosmological constant is at infinite Weyl distance in the space of all $AdS$ metrics.
This observation leads us to the following Ricci flow conjecture:

\vskip0.4cm

  {\bf Conjecture A:} {\sl Consider quantum gravity on a family of background metrics $g_{\mu\nu}(t)$ satisfying the Ricci flow equation \eqref{rf}. There exists an infinite tower of states which become massless when following the flow towards a fixed point $\hat g_{\mu\nu}$ at infinite distance.}
 
\vskip0.3cm
In general, the metric distance $\Delta_g(t,\hat t)$  along the Ricci flow between $t$ and the fix-point $\hat t$ is a function of $t$ and $\hat t$, which has to be determined for a given background metric.  The associated tower
of states is supposed to scale as

 \be
m(\hat t) \sim m(t) e^{-\alpha | {\Delta_g }(t,\hat t)|}\;.
\label{dscricci}
\ee
Typically for many examples the fix-point $\hat t$ will be either at $\hat t=+\infty$ or $\hat t=-\infty$.
Note  the tower of states can only appear when following the Ricci flow to the fixed point. 
The theory at the fixed point,  like e.g. Minkowski space, does not necessarily contain an infinite tower of massless states, but it is typically part of a larger (higher dimensional) space-time.
However in case of infinite distance, the fixed point theory can never be reached, which means that the transition from the theories along the flow to the fixed point theory is discontinuous.

\vskip0.3cm
Let us check this conjecture
for the simple cases of Einstein spaces and in particular for (A)dS  with non-zero cosmological constant.
For that purpose
let us calculate the evolution of the scalar curvature. It is easy to verify that, using (\ref{rf}) and the definition of the scalar curvature,  the latter evolves as 
\begin{eqnarray}
\frac{\partial R}{\partial t}=\nabla^2 R+2R^{\mu\nu}R_{\mu\nu}. \label{eqs}
\end{eqnarray}
For an AdS or dS space and more general for a $d$-dimensional Einstein space with 
\begin{eqnarray}
R_{\mu \nu}=\Lambda g_{\mu\nu}, \label{ein}
\end{eqnarray}
 we find that Eq.(\ref{eqs}) leads to the following simple Ricci flow equation for the cosmological constant:
\begin{eqnarray}
\frac{\d \Lambda}{\d \t}=2 \,\Lambda^2. \label{ll}
\end{eqnarray}
It is obvious that the fixed point of the flow equation for the cosmological constant is at $\Lambda=0$.
The explicit  solution to the above equation is  
\begin{eqnarray}\label{soll}
\Lambda(\t)=\frac{\Lambda_0}{1-2\Lambda_i(\t-\t_i)},
\end{eqnarray}
where $\Lambda_i$ is the cosmological constant at some initial time $\t_i$. 
\begin{eqnarray}
\Lambda_i=\Lambda(\t_i).
\end{eqnarray}

We see that $|\Lambda(\t)|$, depending on its sign, approaches zero at  $\t\to \infty$ or $t\to -\infty$. 
In particular for an initial AdS space with negative $\Lambda_i<0$,  the Ricci flow ends at flat space at $\t\to \infty$  (fixed point of the Ricci flow) corresponding to an immortal Ricci soliton. 
  If on the other side, $\Lambda_i>0$ is positive, the cosmological constant approaches zero and an initial dS space ends at a flat space 
at $\t\to -\infty$ corresponding to a ancient Ricci soliton. 

Note that,  using the explicit $AdS$ metric,
 the general
Ricci flow equation (\ref{rf}) also immediately leads to the Ricci flow equation (\ref{ll}) for the cosmological constant.

On the other hand,  for the path derived from the canonical Weyl rescalings (\ref{weytra1}),  the cosmological constant evolves as a function of $\tau$ in the following way:
\be
\Lambda(\tau) =  e^{-2\tau}\Lambda_0 \;. \label{ll0}
\ee
Comparing then (\ref{soll}) and (\ref{ll0}), we find that the parameter $t$ in the Ricci flow and the parameter $\tau$ in the canonical Weyl rescaling are both affine parameters of the same path in field space if they are related by 
the following reparametrization:
\begin{eqnarray}
\tau=\frac{1}{2}\ln\Big(1-2\Lambda_i(\t-\t_0)\Big)\, .\label{reltaut}
\end{eqnarray}
Since we are working in the regime $\tau\geq 0$, $\t-\t_i$ must be positive for AdS space and it must be negative for dS.\footnote{Formally we could also consider the "dual" limit $(t-t_0)\rightarrow 1/2\Lambda_0$ with $\tau
\rightarrow-\infty$,
but this would imply an infinite cosmological constant and infinite curvature.}

It should be noted that the relation (\ref{reltaut}) is generic for Einstein spaces. Indeed, let us consider an Einstein space with 
\begin{eqnarray}
R_{0\mu\nu}=\Lambda_i \overline {g}_{0\mu\nu}.
\end{eqnarray}
Then the solution to the Ricci flow equation (\ref{rf}) is 
\begin{eqnarray}
g_{\mu\nu}(t)=\Big(1-2\Lambda_i t\Big)\overline g_{0\mu\nu}.  \label{ei}
\end{eqnarray}
In other words, it is just a Weyl rescaling of the initial metric. If the latter is parametrized as in (\ref{weytra1}), then (\ref{reltaut}) follows.

 We may also explicitly calculate the distance in field space of the AdS or dS, or more general of the Einstein space (\ref{ein}) to the flat space at $\t\to \infty$ or $\t\to-\infty$
 along the Ricci glow. Using  the metric distance formula (\ref{dismetr})
for the Ricci solitons with cosmological constant satisfying (\ref{ll}), we find that  the metric distance $\Delta_g$ in field space is 
\begin{eqnarray}
\Delta_g&=&2\sqrt{d}|\Lambda_i| c\int_{\t_i}^{t_f}\!\!\!\!\frac{\d t}{1-2\Lambda_i(\t-\t_i)}\nonumber \\
&=& c\sqrt{d}\, \ln\Big(1-2 \Lambda_i(\t-\t_i)\Big)\Big|_{\t_i}^{t}
\, . \label{drr}
\end{eqnarray}
Using the reparametrization eq.(\ref{reltaut}) between $t$ and $\tau$, this is nothing else than the distance with respect to Weyl rescalings,
$
\Delta_g=2c\sqrt{d}\,  \tau$, 
and hence we see  that the distances with respect to Weyl rescalings and along the Ricci flow agree with each other, namely they are both proportional 
to the  Weyl rescaling parameter $\tau$.
 In fact, here the Ricci flow and the Weyl rescaling are just reparametriziations of the same path. So this follows of course from the reparametrization invariance of the definition of $\Delta_g$.
 In particular, the distance of an initial AdS space to the  flat fixed point space is logarithmically divergent in Ricci flow parameter $t$ or linearly divergent in Weyl rescaling parameter $\tau$ and therefore
 at infinite distance.

 One may wonder if the above analysis is a peculiarity of Einstein manifolds. In the next section about strings we will consider another example of a non-Einstein homogeneous space,
 namely the so-called Nilmanifold of negative curvature, which will again flow to vanishing curvature.
  On the other hand for spaces $M_d$ with positive curvature, the manifold is contracting, and the point where ${\partial \over\partial\tau}g_{\mu\nu}(t)$ is zero is an unstable fixed point, which might be nevertheless at infinite distance away
from any other point within the family of metrics $g_{\mu\nu}(t)$. 

Let us consider also the case in which the Riemann space may get both signs of the Ricci scalar $R$ in its moduli space under the Ricci flow. 
%Let us note here that from the Ricci-flow equation we get for the volume of the manifold ${\cal M}$ 
%\begin{eqnarray}
% \frac{\partial} {\partial t}\int_{\cal M} \sqrt{g}\, d^n x=-\int_{\cal M} R \,\sqrt{g}\, d^n x.
% \end{eqnarray} 
%Therefore, the flow cannot connect negative to positive scalar curvature  since such transition will pass through $R=0$ which is a fixed point of the volume flow.   Only if this point is at infinite distance we will have the infinite tower of massless states. 
Such a case is provided for example by the  squashed  $S^3$
(although it is not meant to be a solution in string theory).
There are regions in the moduli space where  $R$ is positive, negative or zero. Indeed, the scalar curvature for the squashed $S^3$ with metric 
\begin{eqnarray}
ds^2=\lambda_1^2(t) \sigma_1^2+\lambda_2^2(t)\sigma_2^2+\lambda_3^2(t)\sigma_3^2 \,,
\end{eqnarray}
where $\sigma_i ~(i=1,2,3)$ are the $SU(2)$ left-invariant  one-form, takes the form
\begin{eqnarray}
R=\frac{1}{2\lambda_1^2\lambda_2^2\lambda_3^2}\Big(2 \lambda_1^2 \lambda_2^2+2\lambda_1^2 \lambda_3^2+2\lambda_2^2\lambda_3^2-\lambda_1^4-\lambda_2^4-\lambda_3^2\Big) \,.
\end{eqnarray}
In addition, the Ricci flow equations are explicitly 
 written as 
\begin{eqnarray}
\partial_t\lambda_1^2=-\frac{\lambda_1^4-(\lambda_2^2-\lambda_3^2)^2}{\lambda_2^2\lambda_3^2}, ~~~\partial_t\lambda_2^2=-\frac{\lambda_2^4-(\lambda_1^2-\lambda_3^2)^2}{\lambda_1^2\lambda_3^2},~~~\partial_t\lambda_3^2=-\frac{\lambda_3^4-(\lambda_1^2-\lambda_2^2)^2}{\lambda_1^2\lambda_2^2} \,.
\end{eqnarray}
By inspecting the above equations one may infer that starting for example with an initial metric with positive scalar  curvature, the flow will always be along metrics with positive scalar curvature. In other words, crossing $R=0$ is not possible by the flow. In fact, this is not restricted only to the squashed sphere but is a general result \cite{Knopf}.  
Indeed, 
it is easy to see that even if we start with $\lambda_1=\lambda_2$, the point 
$\lambda_3=2\lambda_1$, which correspond to $R=0$, can never be reached. This shows that although there are points in moduli space that have $R=0$, one should check that such points are reachable under the Ricci flow. In fact, it is known that 
the Ricci flow not only of the squashed $S^3$ but of any homogeneous metric on $SU(2)$  shrinks to a point in finite time and  asymptotes  to the round $S^3$ \cite{Knopf}.  

 This is also consistent with  the $SU(2)_k$ WZW model which has positive curvature in the entire moduli space, given by the positive level k of the Kac-Moody
  algebra. For large positive k the curvature become zero, and this is a fixed point of the backwards Ricci flow, which is  at infinite distance.
   One can go to negative levels k and indeed then the $SU(2)_k$ WZW describes a hyperbolic sphere with negative curvature. However the transition from positive k
  (i.e.~positive curvature) to negative k (i.e.~negative curvature) occurs in the non-geometric regime of the WZW model.

 \subsection*{String theory (quantum gravity) realization}

At the beginning of this section  let us address the question, why the Ricci flow is a useful concept  and provides sensible answers concerning the large distance discussion in string theory and in quantum gravity.
For the moment let us consider purely geometrical backgrounds in string theory, given in terms of a background metric $g_{\mu\nu}(t)$, where for the moment we do not add
 any other background fields like 
 the dilaton $\phi$. Clearly, the Ricci flow is only non-trivial for curved, d-dimensional  manifolds $M_t^d$ with non-vanishing Ricci tensor. In case the Ricci flow has a fixed point at infinite distance, then the scalar curvature flows to zero  and the  Ricci flow conjecture ``correctly” states that these manifolds must be supplemented by an infinite tower of states in order to be consistent in string theory or quantum gravity. In a usual string theory embedding
  the curved manifolds $M_t^d$ are only part of the total 10-dimensional space time (see also below), and the change of the flow parameter $t$ of $M_t^d$ is correlated
 to  a simultaneous change of another parameter in the remaining part of the total space.  
 
 As we will discuss, the formulation of the infinite distance conjecture using the Ricci flow provides a very good and geometric definition of the distance in the space of background metrics
 in terms of the scalar curvature of  $M_t^d$.
 Even more, this definition of the distance can be nicely generalized for the case of adding other background fields, like the dilaton, via use of generalized gradient flows and their associated
 entropy functionals.
A strong argument, why the distance along the Ricci flow is "superior" to the original generalized distance conjecture,
 is provided by backgrounds,  which are consistent in string theory/quantum gravity without the necessity to add an additional tower of massless states.
Consider 
 for example  a consistent string background of the form
\begin{equation}\label{eq:S1comp}
M^{10}={\mathbb R}^{1,8} \times S^1 \, ,
\end{equation}
where $S^1$ is a circle.
Now sending  the radius $R$ of $S^1$ to very large values does not lead to a tower of states on $M^{10}$. This theory is complete in quantum gravity as it stands.
Only in the effective  ${\mathbb R}^{1,8}$ theory there is a tower of massless states in the limit of $R\rightarrow\infty$,
because the distance in the scalar fields space of the modulus $R$ is infinite in the decompactification limit.
However applying the original infinite distance arguments to the metric of ${\mathbb R}^{1,8} \times S^1$
 would lead to a misleading result, it would predict an infinite tower of states on the full space $M^{10}$.
On the other hand, $    {\mathbb R}^{1,8} \times S^1    $ being flat,  it does not flow under the Ricci-flow, and therefore the Ricci-flow conjecture applied to $M^{10}$ 
leads to a correct result.\footnote{We thank E. Palti for discussion
on this point.}

 Furthermore using the Ricci flow, one realizes that
 the infinite distance conjecture has a close connection to renormalization group (RG) flow in the underlying 2-dimensional string $\sigma$ model:
 Being non Ricci-flat, the curved manifolds $M_t^d$ themselves do not satisfy the 2-dimensional $\beta$-function equations for the metric. (Of course the total 10-dimensional space has to satisfy all 2-dimensional
 $\beta$-function equations.) Therefore the Ricci flow from curved spaces to their fixpoints
 is equivalent  to consider  the renormalization group (RG) flow 
 from curved  background spaces $M_t^d$ towards the solutions of the $\beta$-function equations, which 
are just the CFT backgrounds in string theory with vanishing two-dimensional
 $\beta$-function.\footnote{See   e.g.    
  \cite{Bakas:2005kv,Bakas:2007qm}      for a more detailed discussion about the relation between the geometric Ricci flow and the 2d RG flow.
Furthermore the connection between the swampland conjecture and the RG in energy was also discussed recently in  \cite{Gomez:2019ltc}.}

E.g., the AdS space flows to flat space, which is at infinite distance and obviously solves the metric beta function equations.
 In general, the  CFT fixed point $\hat M^d$ might be at infinite distance, and then the infinite distance conjecture is stating that the background $M_t^d$ in the neighbourhood of the fixed point must be always
accompanied by an infinite tower of states. In case of infinite distance the CFT fixed point can never be reached, which means that the transition to the CFT fixed point is discontinuous.
However the CFT fixed point also can be at finite distance and then there is no tower of massless states. In this case the transition to the CFT fixed point is continuous.
In the next section, we will provide an example of a flow to a CFT fixed point, which is at finite distance.

\vskip0.4cm

In case the fixed point is at infinite distance one has to ask  what is the origin of the massless tower of states in string theory or in quantum gravity.
One typical way to answer this question is to 
 supplement the  backgrounds $M_t$ (like AdS) with an extra space factor $K_{t}$ (like a sphere) 
 which, when $M_t$ flows
 to flat space, typically flows to flat space, too. 
Then the KK modes of $K_{t}$  are the origin of the massless tower in the effective field theory of $M_t$. 
So, 
typical string theory or M-theory examples, where the Ricci-flow argument applies, are of the form
%\vskip0.1cm
\begin{eqnarray}
%\boxed
%{
M_t\times K_{t}\, .
\end{eqnarray}
%\vskip0.1cm
The massless tower can however also have a different origin. The associated states can also correspond to tensionless strings with 
light string excitations (Regge modes), which become massless in the infinite distance limit, or also to wrapped tensionless strings and branes.

\vskip0.4cm
Another related question is if one can describe the infinite distance arguments via Ricci flow for the moduli space of Calabi-Yau manifolds.
As already said before, for Ricci flat metrics like string 
compactifications on 
 a Ricci-flat Calabi-Yau manifold or on a circle,
there is no Ricci flow along the marginal directions at all. 
This is in agreement with our conjecture as the full ten-dimensional theory is consistent at all points in the Calabi-Yau moduli space.
Only in an effective lower-dimensional description, where the internal Calabi-Yau manifold is not part of the (lower-dimensional) background metric, there is an infinite tower of massless states when going to an inifinite distance point in moduli space. 

One way to bypass the situation 
that the Ricci-flow does not exist for marginal directions of Ricci-flat spaces, let us 
consider instead manifolds, where the Ricci-flat Calabi-Yau space is non-trivially fibred over some base, such that the total space has non-zero curvature. As we will show in the
following, now the Calabi-Yau (K\"ahler) moduli will start to flow, and in this way one can  test the Ricci-flow conjecture also for Calabi-Yau spaces. 
The prototype example of this kind is
a d-dimensional spacetime of the form $M_d=\tilde M^{d-3}\times N$, where $N$ is the  Nilmanifold, namely the  3-dimensional twisted torus with metric 
\begin{eqnarray}
 ds^2=\!a^{2}(dx_1^2\!+\!dx_2^2)\!+\!b^{2}(dx_3\!+\!x_1dx_2)^2\!\, .
 \end{eqnarray} 
 The Nilmanifold has negative curvature and can be viewed as a circle fibration  over a torus.
 We will be mainly interested in the circle $S^1$ in the $x_1$ direction,  whose radius, being proportional to $r_1=a$, stays constant under the fibration.
 As we will see
 the Ricci-flow is non-trivial, it leads to a flow with respect to the radius $r_1$, which ends at the infinite radius fixed point, where $a\rightarrow\infty$ and the curvature of the Nilmanifold is vanishing.
Specifically, for $a=a(t)$ and $b=b(t)$, the Ricci flow (\ref{rf}) is determined by the equations
\begin{eqnarray}
x'(t)&=& \frac{y(t)}{x(t)},\\
y'(t)&=&-\frac{y(t)^2}{x(t)^2}, ~~~~x(t)=a(t)^2, ~~y(t)=b(t)^2,
\end{eqnarray}
where prime denotes derivative with respect to $t$. 
From the above equations we find that 
\begin{eqnarray}
x\,y=C_0. \label{xy}
\end{eqnarray}
In addition,  the Ricci scalar 
\begin{eqnarray}
 R(t)=-\frac{y(t)}{2x(t)^2}, \label{rr}
 \end{eqnarray} 
 satisfies the equation
 \begin{eqnarray}
 R'=6R^2,
 \end{eqnarray}
from where we get
\begin{eqnarray}
 R(t)=\frac{R_i}{1-6R_it} \quad{\rm with}\quad R_i<0\, .
 \end{eqnarray} 
 Since we do not want to enter the regime of large curvature, we restrict the flow parameter to the regime $t\geq 0$. At the end of the discussion we will also comment on the large curvature limit where $t\rightarrow 1/6R_i$.
 Therefore, from Eqs.(\ref{xy}) and (\ref{rr}) we get  
 \begin{eqnarray}
 a^2&=&\left(-\frac{C_0}{R_i}\Big(1-6 R_i t\Big)\right)^{1/3}\nonumber \\
 &=&a_0^2\Big(1-6 R_i t\Big)^{1/3},\\
 b^2&=&\frac{C_0}{\left(-\frac{C_0}{R_i}\Big(1-6 R_i t\Big)\right)^{1/3}}\nonumber \\
 &=&\frac{b_0^2}{\Big(1-6 R_i t\Big)^{1/3}}, ~~~b_0^2 a_0^2=C_0.
 \end{eqnarray}
Notice, that in this case, the Ricci flow is not simply a Weyl rescaling of the metric since  the Nilmanifold is not an Einstein space. 
 On the other hand, under  the Weyl rescaling (\ref{weytra1}) on the initial metric 
 with scalar curvature $R_i$, the latter transforms as 
 \begin{eqnarray}
 \hat R(t)=e^{-2\tau}R_i, 
 \end{eqnarray}
from where we find  that\footnote{ Note that we compare here the transformation of the curvature scalar under rescalings with that of the curvature under the Ricci flow since the metric undergoes an anisotropic dilation.  We could of course compare other geometric quantities like the volume for example, which is also rescaled under the Ricci flow. }
\begin{eqnarray}
 \tau=\frac{1}{2}\ln \Big(1-6 R_i t\Big), ~~~\tau\geq 0.
 \end{eqnarray} 
 Then, from Eq.(\ref{dismetr}) we may  calculate the distance $\Delta_g$, which turns out to be 
 \begin{eqnarray}
 \Delta_g\simeq \sqrt{3} C_0\int_{0}^{t_f}\frac{dt}{a^3}=\frac{1}{2\sqrt{3}}
 \ln\Big(1-6R_it\Big)\Big|_{0}^{t_f}
 =\frac{1}{\sqrt{3}}\tau\, .
 \end{eqnarray}
Note that points at infinite distance are at $\tau\to \infty$, which corresponds to $t\to \infty$ 
%and $t\to 1/6R_0$. 
At this point we have the behaviour 
\begin{eqnarray}
&& t\to \infty, ~~~~~~a\to \infty, ~~~b\to 0\, .
%&& t\to \frac{1}{6R_0}, ~~\,~a\to 0, ~~~~b\to \infty.
\end{eqnarray} 
Therefore, we see that  at $t\to \infty$  the radius $r_1$ of the circle in the first direction becomes very large. The associated 
KK states with masses $m_{KK}\sim 1/a=1/r_1 $ 
become light and build an infinite tower massless states.

Here we can also ask what is possibly happening, if one takes the dual limit with $\tau\rightarrow -\infty$, leading to 
 $t\to 1/6R_i$, $a\to 0$ and $b\to \infty$. This limit 
 corresponds to vanishing radius $r_1\rightarrow 0$.
 The corresponding light states could be identified with the string winding modes, i.e. fundamental strings wrapped around the circle in the first direction. 
 They have masses of the order $m_{F1}\sim a=r_1 $.
But  this limit is not controllable, it corresponds to infinite curvature of the Nilmanifold and is also not
   a fixed point of the Ricci-flow. This is in agreement with the Ricci-flow conjecture which states that the massless tower  of states 
  appears in the controllable regime of general relativity at the fixed point of the Ricci-flow at weak curvature.

We can also try identify the $x_1$ direction with the $S^1$ of M-theory.
From the ten-dimensional point of view the KK states, namely the D0-branes of the IIA string theory  have masses 
\begin{eqnarray}
m_{D0}\sim a^{-9/8}\sim e^{-\frac{3}{4}\tau}\, .
\end{eqnarray}
In the limit $\tau,t\to \infty$ with $r_1\rightarrow \infty$ and with zero curvature they become light and lead to an infinite tower of states.

Again we can look at the dual limit with $\tau\rightarrow -\infty$ in M-theory with vanishing radius $r_1\rightarrow 0$.
 In M-theory
there are now light states, which are dual to the D0-branes, namely they
correspond to IIA D6-branes wrapping the $S^1$. From the ten-dimensional point of view, the corresponding tensions of the light D6-branes are then
 \begin{eqnarray}
    m_{D6}\sim a^{9/8}\sim e^{\frac{3}{4}\tau}.
    \end{eqnarray} 
   Being not particles,  it is however not clear in what sense these wrapped branes correspond to a tower of massless states. 
   Again this is in agreement with the Ricci-flow conjecture.

\vskip0.9cm
Alternatively one could consider quantum gravity or strings on Calabi-Yau spaces $\tilde X^6$ equipped with a non-Ricci flat metric. Here one again expects a Ricci-flow of the complex metric $\tilde X^6$
towards the Ricci flat Calabi-Yau metric of $ X^6$, and the question is, if the fixed point of this flow is again accompanied by an infinite tower of massless states. Actually string compactifications on
\begin{equation}
{\mathbb R}^{1,3}\times \tilde X^6\, ,
\end{equation}
were recently considered in \cite{Larfors:2018nce}, where the associated 
$SU(3)$ group structure of $\tilde X^6$ is non-trivial and is not the same as the $SU(3)$ group structure of its Ricci-flat relative $X^6$.
Since the 4D space $M_4$ is flat Minkowski space, one does not expect an infinite tower of states emerging at the endpoint of this flow. 
However this a priori is not an argument against our Ricci-flow conjecture, since  
these SU(3) structures of $\tilde X^6$    \cite{Strominger:1986uh,Hull:1986kz,LopesCardoso:2002vpf}  
have a non- vanishing and non-closed three-form flux which needs to be supported by source terms in the associated Bianchi identity. 
Theses fluxes and the additional sources most likely do not allow a  flow towards the Ricci-flat Calabi-Yau space, and hence the Ricci-flow argument cannot be invalidated by these spaces.

\subsection{Entropy functionals and generalized distances}

Let us recall that the Gibbons-Hawking entropy \cite{Gibbons:1977mu} is proportional to the area of the event horizon, namely
\begin{equation}
{\cal S}_{GH}=1/\Lambda \,
\end{equation}
Subsequently ${\cal S}_{GH}$ was interpreted as the dimension of the Hilbert space ${\cal H}$ in an observer's causal domain \cite{Banks:2000fe,Witten:2001kn}:
\begin{equation}
{\rm dim}~{\cal H}=e^{1/\Lambda}\, .
\end{equation}
In the limit $\Lambda\rightarrow 0$ this quantity becomes infinite. This observation was used in \cite{Ooguri:2018wrx} to relate the refined de Sitter conjecture to
the entropy of tower of massless states, which appear  at infinite distance in the limit $\Lambda\rightarrow 0$. 
In this section we want to describe the distance via the entropy functionals of Perelman, which  played an important role
for the proof of the Poincare conjecture by Perelman.

As explained before (see eq.(\ref{lambdadistance}))  the distance $\Delta$ is related to the (A)dS cosmological constant $\Lambda$ in the following way:
\begin{equation}
\Delta\simeq-\log |\Lambda|\,  \label{dffg}.
\end{equation}
For de Sitter space with $\Lambda>0$, one gets that 
\begin{equation}
\Delta\simeq\log {\cal S}_{GH}\, .
\end{equation}
Therefore the entropy becomes large in the large distance limit, which leads a large tower of light states.

On the other hand, $\Lambda$ is part of the effective string action, which is given as
\begin{equation}
S_{\rm eff}={1\over 2}\int d^dx\sqrt{-g}\biggl( R-(\nabla \phi)^2+ \Lambda\biggr)\, .
\end{equation}
Now, trading $\Lambda$ by the scalar curvature plus the dilaton action, one could propose that the distance $\Delta$ is given as
\begin{equation}
\Delta\simeq \log \frac{S_{\rm eff}}{V_M}\, ,
\end{equation}
where $V_M$ is the d-dimensional volume. 
However as we will discuss in the following, the more useful definition of the distance is given in terms of generalized entropy functionals of the relevant gradient flow.

\subsubsection{Distance for the Ricci flow in terms of the scalar curvature $R$}

As discussed above, only considering the geometric distance \eqref{dismetr} on the space of background metrics can lead to misleading results.
There are simple examples of families of metrics which can be all used as backgrounds in a consistent quantum gravity theory without including an additional tower of light states, even in the infinite distance limit.
For these examples it is only upon compactification that such a tower appears.
However, taking into account only distances which are measured along the Ricci flow can cure this ambiguity.
For this reason we would now like to use our previous results to go away from the geometric definition \eqref{dismetr} and to propose alternative, more abstract distance measures.
This will also allow us to accommodate for additional fields like the dilaton.

\vskip0.4cm
Let us first consider
pure Einstein gravity and  the corresponding Ricci-flow. Here we like to formulate the  following  distance conjecture for the Ricci flow:
\vskip0.3cm

{\bf Conjecture B1:} {\sl For a d-dimensional Riemannian manifold, the  distance $\Delta_R$ in the field space of the background metrics along the  Ricci-flow 
is determined by the scalar curvature $R(g)$.
At ${ R}=0$ there is an infinite tower of additional massless states in quantum gravity.}

\vskip0.3cm
Furthermore 
we propose an alternative definition for the distance 
to be used in the mass formula \eqref{dscricci}, along the Ricci flow between an initial flow parameter $t_i$ and a final flow parameter $t_f$:
\begin{equation}
\Delta_R\simeq \log \Bigg({R_i\over R_f}\Biggr)\,.\label{resultR}
\end{equation}
Here $R_i$ and $R_f$ are the corresponding initial and final values of the scalar curvature.
Note that we sometimes follow the flow backwards, e.g. $t_f < t_i$; this is  related to the additional sign in eq.(\ref{additionalsign}).
In the vicinity of the fixed point $R_f=0$ and assuming that $R_i$ is non-vanishing we can write this formula in the following simpler way:
\begin{equation}
\Delta_R\simeq \log  R\, ,\label{resultRa}
\end{equation}
Using this distance formula one expects a tower of massless states which scale in the infinite distance limit $\Delta_R\rightarrow\infty$, $R\rightarrow 0$, as 
\be
m \sim M_p e^{-\alpha |\Delta_R|}\simeq  \bigl( R\bigr)^\alpha\;,
\label{RicciRdisc}
\ee
We see that in general the flat space limit should be accompanied by an infinite number of states. The flat space itself is infinitely far away from the 
curved manifolds along the flow, which means that the flat space never be reached. In other words the transition to flat space is discontinuous.
In a certain sense, the associated entropy ${\cal S}\simeq e^{\Delta_R}$ of space-time in the flat space limit becomes
infinite.
In fact, that flat space has infinite entropy was already pointed out in    \cite{Dvali:2015rea}   in the context of black hole entropy considerations.

\vskip0.3cm
Let us try to make contact between  the definition of the  distance $\Delta_R$ in (\ref{resultR}), which contains the curvature $R$, and the original metric distance $\Delta_g$ given in eq.(\ref{dismetr}).
Assume there is a Ricci flow such that
\begin{equation}\label{eq:riccisquareass}
R_{\mu\nu} R^{\mu\nu} = \frac{k}{2} R^2 \,,
\end{equation}
for some constant $k$ and with $R$ the scalar curvature.
Moreover, we also assume that $R$ is constant over the manifold, i.e.~it is only a function of the flow parameter $t$. 
This class of manifolds is a slight generalization of Einstein Manifolds and includes e.g. also the Nilmanifold.
Then \eqref{eqs} becomes
\begin{equation}
\frac{d R}{d t} = k R^2 \,,
\end{equation}
which can be straight forwardly integrated such that
\begin{equation}
R(t) = \frac{R_i}{1- k R_i t} \,.
\end{equation}
We can now use \eqref{eq:riccisquareass} and the fact that $R$ depends only on $t$ to compute the distance $\Delta_g$,
\begin{equation}\label{additionalsign}
\Delta_g = c \sqrt{2 k} \int_{t_i}^{t_f} \! d t \, \left| {  R(t)}\right| = 
c \sqrt{\frac{2}{k}}\int_{t_i}^{t_f} \! d t \, \left| {R'(t)\over  R(t)}\right| = 
c \sqrt{\frac{2}{k}} \mathrm{sgn}(R_i) \log\left(\frac{R_f}{R_i}\right) \,.
\end{equation}
For Einstein manifolds one has $k = \frac{2}{d}$, and
the three-dimensional Nilmanifold  satisfies \eqref{eq:riccisquareass} with $k = 6$, which reproduces the discussion there.
Hence for this class of manifolds one precisely obtains the advertized result in eq.(\ref{resultR}).

\vskip0.3cm
What about spaces with non-constant curvature $R(t,x^\mu)$? For this spaces the distance now is assumed to take the form 
\be
\Delta_R \simeq \log \Bigl({\bar R(t_i)\over \bar R(t_f)}\Bigr)\,,
\label{resulta}
\ee
where $\bar R(t)$ is the {\sl mean curvature}.

\vskip0.1cm

In fact, for a flow which ends up at the fixed point, which is flat space with $\bar R(t_f)\equiv R(t_f)=0$, but where $\bar R(t_i)\neq0$, we can still write the distance near the fixed point approximately as
\be
\Delta_R \simeq \log  \bar R(t)\,,
\label{result}
\ee
and the distance to reach the fixed point at $R(t_f)=0$ is indeed infinite.

%\vskip0.3cm
%
%A subtle exception is where the fixed point is flat space with $ R(t_f)=0$, but the starting space is non-flat, but nevertheless has vanishing mean curvature $\bar R(t_i)=0$.
%In this case, using (\ref{resulta}) the distance is finite. 
%An interesting example of this kind involves 
%a 2d space with metric  
%%in \eqref{eq:so2metric},
%\begin{equation}
%d s^2 = e^{\phi(r,t)} \left(d r^2 + r^2 d \theta^2 \right) \,,
%\end{equation}
%where the Ricci flow equation reads
%\begin{equation}
%\frac{\partial \phi}{\partial t} = e^{-\phi} \left(\frac{\partial^2 \phi}{\partial r^2} + \frac{1}{r} \frac{\partial \phi}{\partial r} \right)\,.
%\end{equation}
%Let us now consider the flow of a metric which deviates from flat Euclidean space only in some compact region.
%To be more specific, let the initial conditions be given by a smooth profile of the form
%\begin{equation}
%\phi(r,0) = \begin{cases}
%g(r) & r < 1 \\
% 0 & r \geq 1
%\end{cases} \,,
%\end{equation}
%such that $g(r)$ has no divergences.
%The corresponding metric is at finite distance from the flat metric  everywhere.
%A simple example for such a function is given by 
%\begin{equation}
%g(r) = 1 - \frac{e^{-\frac{1}{r}}}{e^{-\frac{1}{r}} + e^{-\frac{1}{1-r}}} \,,
%\end{equation}
%which satisfies $g(r\rightarrow0) = 1$ and $g(r\rightarrow1) = 0$.
%A numerical analysis  shows that the Ricci flow with such initial conditions flows for $t\rightarrow \infty$ towards the flat metric.
%However as just said before the corresponding distance is finite and one can indeed show that the average curvature of the space is always zero.

\subsubsection{Distance for the metric-dilaton  flow in terms of the entropy functional ${\cal F}$}

Studying backgrounds which are more complicated than Einstein manifolds, where we have argued that the Ricci-flow gives a good distance measure,  is only possible in the presence of additional fields. 
Hence, one should study more complicated flows and associated distance functionals. This is in general quite difficult, therefore 
 we first want to include only the dilaton field. The relevant quantities, which are monotonic along the flow, are the entropy functionals of Perelman \cite{Perelman:2006un}. 
Note that these are slightly different from the effective string action - see also the comments at the end of this subsection.
Using the notation of  \cite{Perelman:2006un}, Perelman first considers the following entropy functional:
\begin{equation}
{\cal F}(g,f)=\int d^dx\sqrt{-g}\biggl(R_g+(\nabla f)_g^2\biggr)e^{-f}\, .
\end{equation}
Actually, ${\cal F}(g,f)$ and $S_{\rm eff}$ are closely related quantities, and $f$ basically plays the role of the dilaton $\phi$ in string theory, namely $f=2\phi$, with ${\cal F}$ being the string effective action in the string frame.
The string coupling constant is then $g_s^2=e^f$.
Varying ${\cal F}$ while keeping the integral 
$
\int d^dx\sqrt{-g}e^{-f}
$
fixed, 
the combined metric-dilaton gradient flow is then determined by the following two differential equations:
\begin{eqnarray}
{\partial \over\partial t}g_{\mu\nu}(t)&=&-2\biggl(R_{\mu\nu}(t)+\nabla_\mu\nabla_\nu f(t)\biggr) \, ,
\nonumber\\
{\partial \over\partial t}f(t)&=&-R(t)-\Delta f(t)\, .\label{rfdilaton}
\end{eqnarray}
In fact, after appropriate diffeomorphism, the above equations can be written as 
\begin{eqnarray}
{\partial \over\partial t}g_{\mu\nu}(t)&=&-2R_{\mu\nu}(t) \, ,
\nonumber\\
{\partial \over\partial t}f(t)&=&-R(t)-\Delta f(t)+(\nabla f)^2\, .\label{rfdilaton1}
\end{eqnarray}

In analogy to the case of the pure Ricci flow case considered before we now 
formulate  the following second conjecture about the combined metric-dilaton flow:

\vskip0.3cm
{\bf Conjecture B2:} {\sl For a d-dimensional Riemannian manifold, the  distance $\Delta_{\cal F}$ in the background field space  along the  combined dilaton-metric flow
is determined by the entropy functional ${\cal F}(g,f)$.
At ${\cal F}=0$ there is an infinite tower of additional massless states in quantum gravity.}
\vskip0.3cm

Furthermore 
we propose that  the generalized distance between an initial flow parameter $t_i$ and a final flow parameter $t_f$ can be defined as follows:
\begin{equation}
\Delta_{\cal F}\simeq \log \Bigg({{\cal F}_i\over {\cal F}_f}\Biggr)\, ,
\end{equation}
Here ${\cal F}_i$ and ${\cal F}_f$ are the corresponding initial and final values for the entropy functional.
In case ${\cal F}_f\rightarrow 0$ is a fixed point of the flow equations,
the tower of massless states should now scale in the infinite distance limit as 
\be
m \sim M_p e^{-\alpha |\Delta_{\cal F}|}\simeq  \bigl( {\cal F}(\tau)\bigr)^\alpha\;,
\label{RicciRdisc}
\ee

The distance $\Delta_{\cal F}$ describes  the total distance along the combined flow of metric-dilaton field.
 For constant dilaton, where we have only the Ricci flow of the metric, 
$\Delta_{\cal F}$ becomes $\Delta_R\simeq\log R$, which is the correct distance for the metric. 
On the other hand, when one has only a flow of the dilaton, it is easy to see that  $\Delta_{\cal F}  $ 
becomes 
\begin{equation}\label{stringdistance}
\Delta_{\cal F}\simeq\log g_s\, .
\end{equation}
This  is the correct distance for the string coupling constant.\footnote{That this is indeed the correct relation between the distance and the string coupling constant can be seen from the Weyl rescaling, 
which we discuss  in the next chapter of the paper, and where
we get that
$g_s\simeq e^{c\tau}$, where $\tau$ is the Weyl distance.
Another argument, why the distance for the string coupling constant is like in (\ref{stringdistance}) is provided by looking at the effective action for the string coupling constant, i.e. for the dilaton field $\phi=f/2$:
\begin{equation}
{\cal L}_{\rm eff}\simeq (\partial\phi)^2\simeq {(\partial g_s)^2\over g_s^2}\, .
\end{equation}
So the canonically normalized field is indeed $f=2\phi$ and according to the original distance conjecture the distance is
\begin{equation}
\Delta\simeq \phi=2f\simeq \log g_s\, .
\end{equation}}

Before we come to the combined flow,
let us first consider the case, where we keep the metric fixed and only consider a flow in the dilaton field.
The gradient flow then as follows from the variation of  ${\cal F}$  with respect to $f$ turns out to be
\begin{eqnarray}
{\partial \over\partial t}f(t)=-2\Delta f(t)+(\nabla f)^2\, .\label{rfdilatonalone}
\end{eqnarray}
Using this equation, we find that the inverse of the 
string  coupling $g_s=e^{f/2}$ satisfies the heat equation 
(after redefining $t\to 2t$)
\begin{eqnarray}
{\partial \over\partial t}g_s^{-1}(t)=-\Delta g_s^{-1}(t).
\end{eqnarray}
A simple  solution of the heat equation in one spatial dimension is:
\begin{equation}
g_s^{-1}\simeq \,e^{x-t}\, .
\end{equation}
 The fixed point is at $t\rightarrow+\infty$ where $g_s$ becomes very large, i.e. we deal with strong string coupling.
 This fixed point is at infinite distance, and the corresponding entropy functional 
${\cal F}\simeq 4 a^2e^{2x-2t}\simeq g_s^{-2}$ is vanishing at the fixed point. And the corresponding  distance is indeed $\Delta_{\cal F}\simeq\log g_s$.
 Hence we expect a tower of massless states, when we approach the fixed point.
Since this tower of states emerges at strong string coupling, it cannot be given in terms 
of perturbative string states, but the tower of states must be related to non-perturbative states like wrapped branes.  As we will discuss in the next section,
there is a dual infinite distance point with respect to Weyl rescaling at weak string coupling, where the tower of additional states corresponds to the excited string modes.

\vskip0.4cm

Now let us consider a combined flow of the metric-dilaton background.
For a d-dimensional Einstein space in particular, the metric is given by (\ref{ei}) and the solution for $f$ of (\ref{rfdilaton1}) is 
\begin{eqnarray}
g_s(t)^2=e^{f(t)}=\Big(1-2\Lambda_i t\Big)^d.
\end{eqnarray}
Then, the value of ${\cal F}$ turns out to be
\begin{eqnarray}
{\cal F}=\frac{d \, \Lambda_i}{1-2\Lambda_i t} V_{ M}=d\, 
\Lambda(t) V_M ,
\end{eqnarray}
whereas the distance $\Delta_{\cal F}$ is 
\begin{eqnarray}
\Delta_{\cal F}\simeq -\ln |\Lambda|,
\end{eqnarray}
in accordance with (\ref{dffg}). So the infinite fixed point of the gradient flow is again at vanishing curvature and also at large $f(t)$, which means
that the flow ends at very large string coupling $g_s$.
This is a general feature of the flow since combining the two equations of (\ref{rfdilaton}) we find that $e^{-f}\sqrt{g}=c_0$ where $c_0$ is $t$-independent. Therefore, $e^f$ is small when $\sqrt{g}$ is small, i.e. at collapse, where the  curvature is large, and is large  at small curvature.

However, there are also cases where fixed point solutions of the combined gradient flow are at finite distance. Let us consider for example the  $SL(2,\mathbb{R})/U(1)$ CFT with 
metric \cite{sf1,sf2}
\begin{eqnarray}
ds^2&=&k\left\{\frac{1-\lambda}{1+\lambda}\Big(dr^2+\tanh^2r d\theta^2\Big)\right.
\nonumber  \\
&&\left.+\frac{4\lambda}{1-\lambda^2}\Big(\cos \theta dr-\sin\theta \tanh r d\theta\Big)^2\right\}, 
\label{sm0}
\end{eqnarray}
and dilaton 
\begin{eqnarray}
f=-2\ln(\cosh r). \label{sf0}
\end{eqnarray}
It can easily be verified that (\ref{sm0}) and (\ref{sf0})    satisfy (\ref{rfdilaton}) if $\lambda$ is 
\begin{eqnarray}
\lambda=e^{-\frac{4t}{k}}.  
\end{eqnarray}
At the fixed point $t\to \infty$, the solution above approaches 
\begin{eqnarray}
ds^2\approx k\Big(dr^2+\tanh^2r d\theta^2\Big), ~~f=-2\ln(\cosh r),
\end{eqnarray}
 which is just the Witten cigar \cite{Witten:1991yr}. 
The ${\cal F}$-functional on the other side,  
\begin{eqnarray}
{\cal F}\sim\coth\left(\frac{4t}{k}\right),
\end{eqnarray}
is finite as $t\to \infty$  and therefore the Witten cigar is at finite distance from the $\lambda$-deformed background.

\vskip0.4cm

Let us note at the end of this subsection that eqs. (\ref{rfdilaton}) are similar to the two-dimensional  $\sigma$-model  $\beta$-functions  with dilaton $\phi=f/2$, which are written as
\begin{eqnarray}
\beta_{\mu\nu}&=&\ell_s^2 \Big(R_{\mu\nu}+\nabla_\mu \nabla_\nu f\Big), \nonumber \\
\beta_f&=& 2(d-26)+3 \ell_s^2\Big((\nabla f)^2-2 \Delta f-R\Big), \label{beta} 
\end{eqnarray}
where $\ell_s$ is the string scale. 
Comparing eqs.(\ref{rfdilaton}) and (\ref{beta}), we see that 
\begin{eqnarray}
{\partial \over\partial t}g_{\mu\nu}(t)&=&-\frac{2}{\ell_s^2}\beta_{\mu\nu}, \nonumber \\
{\partial \over\partial t}f(t)&=&\frac{\beta_f}{3\ell_s^2}+\frac{2(d-26)}{3\ell_s^2}+\Delta f-(\nabla f)^2. \label{beta2}
\end{eqnarray}

A final comment concerns the functional ${\cal F}$. One could consider instead of ${\cal F}$ the string effective action ${ S}_{\rm eff}$ which is identical to  ${\cal F}$ but without the constraint of fixed $\int d^d x\, \sqrt{g}\,e^{-f} $. In this case, one gets instead 
\begin{eqnarray}
{\partial \over\partial t}g_{\mu\nu}(t)&=&-2\Big(R_{\mu\nu}+\nabla_\mu\nabla_\nu f-\frac{1}{2}g_{\mu\nu}\big[2\Delta f-(\nabla f)^2+R\big]\Big)
, \nonumber \\
{\partial \over\partial t}f(t)&=&-R-2\Delta f+(\nabla f)^2. 
\end{eqnarray}
We see that  when $f$ is constant, the metric satisfies 
\begin{eqnarray}
{\partial \over\partial t}g_{\mu\nu}(t)&=&-2R_{\mu\nu}+R g_{\mu\nu},
\end{eqnarray}
which does not have even short time solutions as  it implies a backward heat equation for the curvature scalar $R$ \cite{hamilton}.  Only the critical points of the flow generated by ${\cal F}$ and ${S}_{\rm eff}$ are the same (provided $\Delta f=(\nabla f)^2$). In other words, it is the 
${\cal F}$ that generates the flow and not the ${S}_{\rm eff}$. The latter cannot even generate a flow and just gives the critical points.  Therefore, it is the functional ${\cal F}$ that is connected with the distance in field space and it is the one that we have considered here.

\vskip0.4cm

\subsubsection{Distance for the Perelman  flow in terms of the entropy functional ${\cal W}$}

In order to treat also the case of collapsing cycles, Perleman \cite{Perelman:2006un} also introduced another entropy functional. 
It substitutes  the effective action
by the following expression
\begin{equation}
{\cal W}(g,f,\lambda)=\int d^dx\sqrt{-g}\biggl( \lambda(R_g+(\nabla f)_g^2)+ f-d\biggr){e^{-f}\over(4\pi\lambda)^{d/2}}\, , \label{WPer}
\end{equation}
under the constraint
\begin{equation}
\int d^dx\sqrt{-g}{e^{-f}\over(4\pi\lambda)^{d/2}}=1\, .
\end{equation}
Notice that we use $\lambda$ instead of $\tau$ which is usually used in the literature as $\tau$ has already been reserved.  
Here $f$ again corresponds to the dilaton and $\lambda$ is a rescaling factor, which is e.g. closely related to the Weyl rescaling. The new term $fe^{-f}$ together with the $\lambda$ factor
have the effect that the flow of the metric tensor is
entirely determined by the Ricci tensor, as it is the case for the original Ricci-flow. In total, the variation of the functional ${\cal W}(g,f,\lambda)$ leads to the following set of flow equations (after an appropriate diffeomorphism \cite{Perelman:2006un}):
\begin{eqnarray}
{\partial \over\partial t}g_{\mu\nu}(t)&=&-2R_{\mu\nu}(t) \, ,
\nonumber\\
{\partial \over\partial t}f(t)&=&-R(t)-\Delta f(t)+|\nabla f(t)|^2+{d\over 2\lambda(t)}\, ,
\nonumber\\
{\partial \over\partial t}\lambda(t)&=&-1\, .
\label{rfdilatonw}
\end{eqnarray}
This eventually leads us to the following third conjecture about the Perelman flow and the generalized distance conjecture.

\vskip0.3cm
{\bf Conjecture B3:} {\sl For a d-dimensional Riemannian manifold, the  distance $\Delta_{ \cal W}$ in the background field space  along the Perelman flow
is determined by the entropy functional ${\cal W}(g,f,\lambda)$.
At ${ \cal W}=0$ there is an infinite tower of additional massless states in quantum gravity.}
\vskip0.3cm

In analogy to the two cases considered before,  we propose that the distance in field space along the Perelman flow can be defined as follows:
\begin{equation}
\Delta_{\cal W}\simeq \log \Bigg({{\cal W}_i\over {\cal W}_f}\Biggr)\, ,
\end{equation}
Here ${\cal W}_i$ and ${\cal W}_f$ are the corresponding initial and final values for the entropy functional.
In case the limit ${\cal W}_f\rightarrow 0$ is a fixed point of the flow equations,
the tower of massless states should now scale in the infinite distance limit as 
\be
m \sim M_p e^{-\alpha | \Delta_{\cal W}|}\simeq  \bigl( {\cal W}(\tau_f)\bigr)^\alpha\;,
\label{RicciRdisc}
\ee

For a d-dimensional Einstein space we find now that 
\begin{eqnarray}
f(t)=\frac{d}{2}\ln\Big(1-2\Lambda_i t\Big)-\frac{d}{2}\ln t,
\end{eqnarray}
and Perelman's entropy functional ${\cal W}$ turns out to be
\begin{eqnarray}
{\cal W}=\frac{d}{2(4\pi)^{d/2}}\Big(\frac{-2\Lambda_i t}{1-2\Lambda_i t}
+\ln(1-2\Lambda_i t)-\ln t-2\Big)V_M.\nonumber 
\end{eqnarray}

 As an explicit example,  let us consider an $SO(2)$ symmetric 2D space with 
\begin{eqnarray}
ds^2=a(t,r)^2\Big(dr^2+r^2 d\theta^2), ~~~f=f(t,r).
\end{eqnarray}
The corresponding equations for $a$ and $f$ are explicitly written as 
\begin{eqnarray}
 \partial_t a&=&\frac{a''}{a^2}-\frac{{a'}^2}{a^3}+\frac{a'}{r a^2},\\
 \partial_t f&=&\frac{\partial_t a^2}{a^2}
-\frac{f''}{\partial_t a^2}+\frac{{f'}^2}{\partial_t a^2}-\frac{f'}{r}-\frac{1}{t}.
 \end{eqnarray} 
 Remarkably, although the above system of equations look quite  complicated, there is a simple solution which is 
 \begin{eqnarray}
 &&ds^2= \frac{1}{e^{4t}+r^2}\Big(dr^2+r^2 d\theta^2\Big), \label{wc}\\
 && f=-\ln\Big(e^{4t}+r^2\Big)-\ln t+4t.\label{wf}
 \end{eqnarray}
 For $t=const.$, one immediately recognizes the Witten cigar solution.
In order to find $\Delta_{\cal W}$, we need  the value of Perelman's entropy 
(\ref{WPer}). The latter, using (\ref{wc}) and (\ref{wf}) turns out to  be
\begin{eqnarray}
{\cal W}=
%-\frac{1}{4t^2}\Big(3+4t+\ln t\Big),
-\frac{1}{4}\Big(1+e^{-4t}r^2\Big)\left\{1+4t+\log t+\log\Big(1+e^{-4t}r^2\Big) \right\},
\end{eqnarray}
so that 
\begin{eqnarray}
\Delta_{\cal W}=-\frac{1}{a} \log|{\cal W}|.
%-\frac{1}{a} \ln\left(\frac{3+4t+\ln t}{4t^2}\right).
\end{eqnarray}
Notice that ${\cal W}\sim - t$ at $t\to \infty$ (at fixed $r$) so that, according to the conjecture B3,   an infinite tower of additional  massless states should appear in the spectrum.

\section{The Weyl distance in effective field theories}

Let us consider a d-dimensional effective action which contains local operators ${\cal O}_i(g_{\mu\nu},x)$  with corresponding coupling constants $g_i$ of the following form:
\begin{equation}
{\cal S}_{\rm eff}\simeq \int d^dx\sqrt {-g(x)} \sum_i{1\over g_i^2}{\cal O}_i(g_{\mu\nu},x)\, .
\end{equation}
In general the   operators ${\cal O}_i(g_{\mu\nu},x)$  will transform
non-trivially under Weyl rescalings (\ref{weytra1}):
 \be
\tilde{\cal O}_i(\tilde g_{\mu\nu},x)= e^{c'\tau} \;,
%\label{weytra}
\ee
The coupling constants $g_i$ are usually inert under the Weyl rescalings of the metric.

However, on the other hand, consider the case where we replace the operators by some constant, non-vanishing vacuum expectation values:
\begin{equation}
{\cal O}_i(g_{\mu\nu},x)\,\rightarrow\, \langle {\cal O}_i(g_{\mu\nu},x)\rangle=\bar {\cal O}_i\, .
\end{equation}
Therefore let us now consider the Weyl flow (or also the Ricci) flow, keeping the  operator ${\cal O}_i(g_{\mu\nu},x)$ constant under the flow.
If we require that  the product ${1\over g_i^2}{\cal O}_i$ still transforms the same under Weyl rescalings as before, we now have to demand
 that 
 the coupling constants
 transform non-trivially under Weyl rescalings as 
 \be
\tilde{g}_i ^2= e^{-c'\tau} g_i^2\,. 
\label{weytra}
\ee
First consider the infinite distance with respect to Weyl rescalings in the direction $\tau\rightarrow +\infty$ for $c'>0$. For $c'<0$, we should consider the
$\tau\rightarrow -\infty$ instead.
It then follows that there exist an associated tower of states with masses $m_i$ being proportional to some power of the parameter $g_i$:
\be
m_i \sim g_i^\alpha \quad{\rm with}\quad \alpha\geq0\;.
\label{gendisca}
\ee
Alternatively we can also consider the flow in the direction $\tau\rightarrow -\infty$.
Now the infinite tower of states should be associated with the inverse  (dual)  coupling constant, i.e. one gets that the mass of the tower scales as
\be
m_i \sim g_i^{-\alpha} \quad{\rm with}\quad \alpha\geq0\;.
\label{gendisce}
\ee
Which of the two directions, $\tau\rightarrow +\infty$ or $\tau\rightarrow -\infty$, leads to an infinite tower of massless states depends on the microscopic details of the
quantum gravity resp. string theory.
In case there is a strong-weak coupling self-duality one expects that both directions lead to an infinite tower of states.

Let us give some examples. First consider the 4-dimensional gauge kinetic term for a $U(1)$ gauge symmetry:
\begin{equation}
{\cal S}_{\rm eff}\simeq \int d^4x\sqrt {-g(x)} {1\over g_{U(1)}^2}F_{\mu\nu}F_{\rho\sigma}g^{\mu\rho}g^{\nu\sigma}\, .\label{u1}
\end{equation}
As it is believed from the weak gravity conjecture, the limit $g_{U(1)}\rightarrow 0$ is accompanied by a tower of infinitely many light states, and hence is at infinite distance.
This behaviour of the is confirmed in string theory by knowing that 
 $g_{U(1)}$ is a field dependent quantity, given in terms of the moduli of the internal, compact space. Then in particular the tower of KK or winding modes provides the infinite distance
 behaviour of  $g_{U(1)}$, as it can be explicitly checked by computing one-loop threshold corrections to the gauge coupling constants.
 
 Let us emphasize that  here we argue that the infinite distance
 behaviour of  $g_{U(1)}$ already entirely follows from the four-dimensional action, namely from the Weyl rescalings (\ref{weytra}) of the four-dimensional metric $g_{\mu\nu}$.
 In fact it is well-known that the 4D $U(1)$ action (\ref{u1}) is invariant under Weyl invariant, and the operator $F^2$ transform under Weyl rescalings as
 \begin{equation}
 \tilde F^2 = e^{-4\tau}F^2\, .
 \end{equation}
 Hence following the previous argument, we should keep the $F^2$ invariant under the Weyl rescaling giving the appropriate Wel weight to the coupling instead. Therefore, we conclude that the $U(1)$ gauge coupling behaves as 
  \be
\tilde{g}_{U(1)} ^2= e^{4\tau} g_{(1)}^2\;.
\ee
Therefore, there is an infinite tower of states with masses that scale as
\be
m_{U(1)} \sim g_{U(1)}^\alpha \quad{\rm with}\quad \alpha\geq0\;,
\label{gendiscb}
\ee
which become massless in the $\tau\rightarrow -\infty$ limit. 
As a remark, one could also could consider the opposite scale limit with $\tau\rightarrow \infty$. The inverse gauge coupling $1/g_{U(1)}$ goes to zero at infinite distance, which corresponds to the magnetic version
of the weak gravity conjecture.

Next consider a four-dimensional effective action with higher curvature term:
\begin{equation}
{\cal S}_{\rm eff}\simeq \int d^4x\sqrt {-g(x)} {1\over g_n^2}R^n\, .\label{rn}
\end{equation}
Here the term $R^n$ denotes all possible combinations of contractions of the Riemann tensor.
Under Weyl rescalings $R^n$ transforms as:
\begin{equation}
 \tilde R^n = e^{-2n\tau}R^n\, .
 \end{equation}
 Hence we require that 
 \be
\tilde{g}_n ^2= e^{2n\tau} g_n^2\;.
\ee
 This implies that at weak coupling  $g_n\rightarrow 0$, there is an infinite tower of states of the form
 \be
m_n \sim g_n^\alpha \quad{\rm with}\quad \alpha\geq0\;.
\label{gendiscc}
\ee
 Therefore the  Weyl distance conjecture implies just from effective field theory considerations that
 the
 limit $g_n\rightarrow 0$ is at infinite distance and therefore it is not possible in string theory or more generally in quantum gravity to discard the infinite higher curvature terms in the effective action.

We can also apply this procedure to Einstein gravity itself. Using the standard Einstein action we see that the Planck mass has to scale under Weyl rescalings as 
 \be
\tilde M_p= e^{-\tau} M_p\;.
\ee
Now consider the infinite distance limit $\tau\rightarrow -\infty$. It follows that large Planck masses, i.e. sending $M_p\rightarrow \infty$, is  accompanied by a tower of massless states:
 \be
m \sim M_p^{-1}\;.
\label{gendiscc}
\ee
Indeed this behaviour is seen via the following (4D) relation
\begin{equation}
M_p=M_s/g_s\, ,\label{mpms}
\end{equation}
and the limit of large $M_p$ is obtained at weak string coupling, i.e. $g_s\rightarrow 0$. 
Since in the string frame the masses of the string excitations scale like $m\simeq g_sM_s\simeq M_s^2/M_p$,
 one indeed expects that the tower of string excitations becomes massless in the tensionless string limit.

There is also a dual infinite distance limit $\tau\rightarrow \infty$. In this limit the Planck mass becomes very small, i.e. $M_p\rightarrow 0$, and we expect a tower of states like
 \be
m \sim M_p\;.
\label{gendiscca}
\ee
Actually the limit $M_p\rightarrow 0$ is the limit of very strong gravitational force, which according to the weak gravity conjecture should be indeed at infinite distance.
In string theory, using the  relation (\ref{mpms}),
 the limit of small $M_p$ is obtained at strong string coupling, i.e. $g_s\rightarrow \infty$. This limit corresponds  to a dilaton flow to large values, i.e. $\phi(t)\rightarrow\infty$, which 
we discussed in the last section on generalized flow equations.

We can also use the relation (\ref{mpms}) to derive the scaling behaviour of the (4D) string coupling constant:
 \be
\tilde g_s= e^{\tau} g_s\;.
\ee
So we see that the weak (strong) coupling limit is at infinite distance, when $\tau\rightarrow-\infty$ ($\tau\rightarrow\infty$)
and the distance is given as $\Delta_{g_s}\simeq \tau\simeq \log g_s$, in  agreement with the distance $\Delta_{\cal F}$ in (\ref{stringdistance}).

\section{The (A)dS distance conjecture from the higher curvature $R^2$ theory}

\subsection*{(A)dS background}

Now consider the 4-dimensional, quadratic $R^2$ action of the form
\begin{equation}
{\cal S}_{\rm eff}\simeq \int d^4x\sqrt {-g(x)} {1\over g_{R^2}}R^2\, ,\label{r2}
\end{equation}
where $R$ is the Ricci scalar.
As it was explained in \cite{Alvarez-Gaume:2015rwa}, in a flat background with $\bar R=0$  the pure $R^2$ action does not propagate a physical mass spin-two graviton state.
However the situation changes if one considers the $R^2$ theory in an (A)dS background with a non-vanishing background metric, i.e. 
\begin{equation}
\bar R\neq0\, .
\end{equation}
In this case 
scale invariance of the $R^2$ action is spontaneously broken, and it follows that the corresponding coupling constant transforms under Weyl rescalings as 
 \be
\tilde{g}_{R^2}^2= e^{-4|\tau|} g_{R^2}^2\;.
\ee
This implies the existence of an infinite tower of states associated to $R^2$ coupling:
 \be
m_ {R^2}\sim |g_{R^2}|^\alpha \quad{\rm with}\quad \alpha\geq0\;.
\label{gendiscd}
\ee

As we will now show, the infinite distance behaviour of the $R^2$ with coupling constant $g_{R^2}$ is equivalent to the infinite distance conjecture (ADS conjecture) of the cosmological constant.
For that  recall that  the pure ${ R}^2$ theory describes a massless scalar field, coupled to standard Einstein gravity with a massless spin-two graviton plus
a cosmological constant $\Lambda$.
To see observe that the action eq.(\ref{r2})
can equivalently be written as 
\begin{eqnarray}
{\cal S}=\int d^4 x \sqrt{-g}\left(\Phi R-\frac{g_{R^2}^2}{4} \Phi^2\right) .
\label{S4}
\end{eqnarray}
The dimension 2 scalar field $\Phi$ plays the role of a Lagrange multiplier and arises in this conformal (Jordan) frame without space-time derivatives. Through its equation of motion
$\Phi$ is proportional  to the background scalar curvature $\bar R$:
\begin{equation}
\Phi={2\over g_{R^2}^2 }\bar R.
\end{equation}
Performing a conformal transformation 
\begin{equation}\label{weyl}
g_{\mu\nu}= {1\over 2}M_p^2\Phi^{-1}\tilde g_{\mu\nu}
\end{equation}
the action (\ref{S4}) can  be written as  
\begin{eqnarray}\label{einst}
{\cal S}=\int d^4 x \sqrt{-\tilde g}M_p^2\left(\frac{1}{2}\tilde R-\frac{3}{4}{\partial_\mu \Phi \partial_\nu \Phi\over\Phi^2} -{g_{R^2}^2\over 16}{M_p^2}\right) \label{E}.
\end{eqnarray}
We see that the cosmological constant is given 
\begin{equation}
\Lambda={g_{R^2}^2\over 16}M_p^2\, .\label{lambdar}
\end{equation}
 We  have in particular:
 \vskip0.2cm
\noindent
(i) de Sitter backgrounds:
\begin{equation}
\bar R>0,\quad g_{R^2}^2>0.
\end{equation}
In this case $\Phi$ is positive and this class of solutions of the $R^2$ theory describes de Sitter like backgrounds with positive cosmological constant $\Lambda$ in
the Einstein frame. Both the signs of the Einstein term as well as of the scalar kinetic term in the Einstein action (\ref{einst}) are such that the spin-2 graviton as well as the scalar
field $\Phi$ are physical, ghost-free degrees of freedom.

 \vskip0.2cm
\noindent
(ii) anti-de Sitter backgrounds:
\begin{equation}
\bar R<0,\quad g_{R^2}^2<0.
\end{equation}
 $\Phi$ is again positive. Now
this class of solutions of the $R^2$ theory describes anti-de Sitter like backgrounds with negative cosmological constant $\Lambda$ in
the Einstein frame. Again the spin two graviton as well as the scalar
field $\Phi$ are physical, ghost-free degrees of freedom.

Since in eq.(\ref{lambdar}) the cosmological constant is given in terms of the coupling constant $g_{R^2}$, it immediately follows that the AdS distance conjecture is directly inherited from the $R^2$ distance conjecture, i.e.
there is a infinite tower of light states in the limit $\Lambda\rightarrow 0$, which scales as
 \be
m_ {\Lambda}\sim |\Lambda|^\alpha \quad{\rm with}\quad \alpha\geq0\;.
\label{gendiscd}
\ee
In other words, the derivation of the AdS distance conjecture in  \cite{Lust:2019zwm} completely agrees with the higher $R^2$ distance conjecture, introduced in this paper.

We can extend the previous discussion by adding the standard Einstein term to the $R^2$ action, which was already considered in \cite{Starobinsky:1979ty}:
\begin{eqnarray}
{\cal S}=\int d^4 x\sqrt{-g}\Big(M_p^2R +{1\over g_{R^2}^2}R^2\Big). \label{W4}
\end{eqnarray}
The ghost-free spectrum of the theory contains one massless spin-2 (the standard graviton), plus a massive scalar, whose mass is determined by the dimensional coupling constant  as
\begin{equation}
m^2 = {g_{R^2}^2\over 6}M_p^2 \, .\label{scalarmass}
\end{equation}
This can be seen by again
introducing a Lagrange multiplier $t$, one can replace the $R^2$ term with,  
$$
2t{ R}-g_{R^2}^2 M_p^2t^2 \, , 
%~\longrightarrow~ t={R \over8 \mu^2}\, ,
$$
which after the $t$ equation of motion, $t={ R} /M_p^2 g_{R^2}^2$, reproduces the initial ${ R}^2$ term. Thus the action can be written in an equivalent way in terms of the scalar field $t$:
\be
{\cal S}=\int d^4x\sqrt{ g } {1\over 2}\left\{M_p^2(1+2t) { R} -M_p^4g_{R^2}^2 t^2   \right\}\,. 
\ee
In this Jordan frame, $t$ looks like a non-propagating field. This however is an illusion of this frame. Performing a rescaling of the metric in order to obtain a canonical Einstein term,
$$
g_{\mu \nu} \rightarrow g_{\mu \nu}\,e^{-\log(1+2t)}\, ,
$$
the action takes the form:
\be
{\cal S}=\int d^4x\sqrt{g } ~\left[M_p^2 { R} -6M_p^2{ \partial_{\mu} t \partial^{\mu} t\over (1+2t)^2}   
-{g_{R^2}^2M_p^4\over 4} { \left( 2t\right)^2  \over (1+2t)^2 } \right]\, ,
\ee
\be
{\cal S}=\int d^4x\sqrt{ g } ~\left[ M_p^2{ R} - \partial_{\mu} \phi \partial^{\mu}\phi -   
{g^2M_p^4\over 4}\left( 1-e^{-\lambda \phi}\right)^2   \right]\, ,
\ee
where in the second equation above we introduced the canonically normalized scalar field $\phi$,
\be
\lambda \phi=\log(1+2t)  ~~{\rm with}~~~ \lambda= \sqrt{2\over 3}(M_p)^{-1}~ .
\ee
Therefore in the  ${R}+{ R}^2$ theory (when written in the Einstein frame), the canonically normalized scalar field $\phi$ possesses the very special Starobisky \cite{Starobinsky:1979ty} potential:  
\be
V=  {g_{R^2}^2\over 8}M_p^4\left( 1-e^{-\lambda \phi}\right)^2 \, .
\ee
Notice that the potential is semi-positive definite as soon as $g_{R^2}^2>0$. 
For large positive values of $\phi$, $V$ asymptotes to a cosmological "constant", while for large negative values the potential grows exponentially. The Minkowski ground state is at $\phi=0$, where the potential vanishes
and the scalar field
acquires a 
  mass as given in eq.(\ref{scalarmass}).
On the other hand for large positive values of $\phi$, the vacuum is described by an approximate de Sitter space with cosmological constant
\be\label{lambdaR}
\Lambda=  {g_{R^2}^2\over 8}M_p^2 \, .
\ee
Therefore the scalar mass  (\ref{scalarmass}) at the Minkowski ground state and the cosmological constant
are related as follows:
\be
\label{sdsdc1}
m ={2\over\sqrt 3}\Lambda^{\frac12} \;.
\ee 
Furthermore we also see that  the asymptotic de Sitter state and the Minkowski ground state are at infinite distance between each other.
Furthermore the limits $\Lambda\rightarrow 0$ as well as the limit of vanising scalar mass $m\rightarrow 0$ are at infinite distance in the four-dimensional metric space.

\section{The spin-two distance conjecture from the higher curvature $W^2$ theory}

\subsection{Flat background}

Let us consider the quadratic Weyl action
\begin{eqnarray}
{\cal S}_{\rm W}=-\frac{1}{2g_{W^2}^2}\int {\rm d}^4 x\sqrt{-g} \, W_{\mu\nu\rho\sigma}
W^{\mu\nu\rho\sigma}. \label{sw}
\end{eqnarray}
Since 
 we have the identity
 \begin{eqnarray}
 W_{\mu\nu\rho\sigma}
W^{\mu\nu\rho\sigma}=R_{\mu\nu\rho\sigma}
R^{\mu\nu\rho\sigma}-2 R_{\mu\nu}R^{\mu\nu}+\frac{1}{3}R^2,
 \end{eqnarray}
we may write 
\begin{eqnarray}
W_{\mu\nu\rho\sigma}
W^{\mu\nu\rho\sigma}=GB+2 R_{\mu\nu}R^{\mu\nu}-\frac{2}{3}R^2,
\end{eqnarray}
where $GB$ is the Gauss-Bonnet scalar
\begin{eqnarray}
GB=R_{\mu\nu\rho\sigma}
R^{\mu\nu\rho\sigma}-4 R_{\mu\nu}R^{\mu\nu}+R^2. 
\end{eqnarray}
Since the integral of the latter is simply $32 \pi^2 \chi$ where $\chi$ is the Euler number and it is the topological invariant, we may express (\ref{sw}) as 
\begin{align}
\!\!{\cal S}_{\rm W}&=-\frac{1}{2g_{W^2}^2}\int {\rm d}^4 x\sqrt{-g}\,  W_{\mu\nu\rho\sigma}
W^{\mu\nu\rho\sigma}\label{wa} \\
&= -\frac{1}{g_{W^2}^2}\int {\rm d}^4 x\sqrt{-g}\,
\Big( R_{\mu\nu}R^{\mu\nu}-\frac{1}{3}R^2\Big) -32\pi^2 \chi. \nonumber 
\end{align}

Under Weyl rescalings the quadratic $W^2$ transform as 
 \be
\tilde{g}_{W^2}^2= e^{-4|\tau|} g_{W^2}^2\;.
\ee
This again implies the existence of an infinite tower of states associated to $W^2$ coupling:
 \be
m_ {W^2}\sim |g_{W^2}|^\alpha \quad{\rm with}\quad \alpha\geq0\;.
\label{gendisce}
\ee

Now let us recall that the Weyl-square action is equivalent to the standard Einstein action plus a massive spin-two field $w_{\mu\nu}$ given by the 
 following
two-derivative action \cite{Bergshoeff:2009hq,Gording:2018not,Ferrara:2018wlb}:
\begin{eqnarray}
{\cal S}_{\rm W}&=&\int_{\cal M} d^4 x\sqrt{-g}\Big{(}M_p^2 R(g)+{2M_p}G_{\mu\nu}(g)w^{\mu\nu}\nonumber\\
&-&M_W^2(w^{\mu\nu}w_{\mu\nu}-w^2)\Big{)}. \label{bimetric}
\end{eqnarray}
Here $G_{\mu\nu}=R_{\mu\nu}-1/2 R\,g_{\mu\nu}$ is the Einstein-tensor constructed from the metric $g_{\mu\nu}$ and the last term is a mass term for the second metric $w_{\mu\nu}$. This action contains a massless spin-two field $g_{\mu\nu}$ 
plus a massive spin-two field $w_{\mu\nu}$. 
Note that the two-derivative kinetic term for $w_{\mu\nu}$ is hidden in the coupling $G_{\mu\nu}(g)w^{\mu\nu}$, 
which can be seen by performing two partial integrations on this term. However after the partial integrations the kinetic term for $w_{\mu\nu}$ has the wrong sign,
i.e. $w_{\mu\nu}$ is a ghost-like field.

The mass of spin-two field $w_{\mu\nu}$ in eq.(\ref{bimetric}) is given as 
\begin{equation}
M_W=g_{W^2}M_p\, .
\end{equation}
Since the coupling constant $g_{W^2}$ exhibits the infinite distance behaviour as shown eq.(\ref{gendisce}), 
it follows that there is a infinite tower of light states in the limit $M_W\rightarrow 0$, which again scales as
 \be
m_{W^2}\sim M_W^\alpha \quad{\rm with}\quad \alpha\geq0\;.
\label{gendiscd}
\ee
Hence this agrees with spin-two swampland conjecture, which was first formulated in \cite{Klaewer:2018yxi}.

\subsection{(A)dS background}

Let us now consider the Weyl-square theory in a curved dS or AdS background and also relate the infinite distance conjecture to the
Higuchi bound for higher-spin fields in de Sitter space.
The equations of motion for the  Weyl theory  are written as 
\begin{eqnarray}\label{weq}
  B_{\mu\nu}=0,
 \end{eqnarray} 
 where   $B_{\mu\nu}$ is the Bach tensor, defined as 
\begin{eqnarray}
 B_{\mu\nu}=\nabla^\rho\nabla_\sigma W^\sigma_{~\mu\rho\nu}+
 \frac{1}{2}R^{\rho\sigma} W_{\rho\mu\sigma\nu}.    \label{bach}
 \end{eqnarray} 
Under conformal transformation of the metric
\begin{eqnarray}
g_{\mu\nu}\to e^{2\tau(x)}g_{\mu\nu},
\end{eqnarray}
it transforms homogeneously as 
\begin{eqnarray}
\widehat B_{\mu\nu}=e^{-2\tau(x)} B_{\mu\nu}. 
\end{eqnarray}
 The Bach tensor is symmetric, traceless due to conformal invariance and due  to diff. invariance, it is also divergence-free
 \begin{eqnarray}
  B^\mu_{~\mu}=0, ~~~\nabla^\mu B_{\mu\nu}=0.
  \end{eqnarray} 
The field equations for Weyl gravity can also be written as 
\begin{eqnarray}
&~&\frac{1}{2}R_{\alpha\beta}R^{\alpha\beta} 
g_{\mu\nu}-2 R^{\alpha\beta}R_{\mu\alpha\beta\nu}+\frac{1}{2}\Box R g_{\mu\nu}\nonumber\\
&-&\Box R_{\mu\nu}-\frac{1}{6}
R^2 g_{\mu\nu}+\frac{2}{3}R R_{\mu\nu}+\frac{1}{3}\nabla_{\mu}\nabla_\nu R=0. \label{fe}
\end{eqnarray}

It is easy to verify that Einstein spaces ($R_{\mu\nu}=\Lambda g_{\mu\nu})$ satisfy the field equation (\ref{fe}). 
Therefore, one such solution is dS space with
\begin{eqnarray}
R_{\mu\nu\rho\sigma}&=&H^2\Big(g_{\mu\rho}g_{\nu\sigma}-g_{\mu\sigma}g_{\nu\rho}\Big), \nonumber \\
R_{\mu\nu}&=&3H^2 g_{\mu\nu}, ~~~R=12H^2.  \label{R3}
\end{eqnarray}
This solution describes a spontaneous 
breaking of local scale invariance of the Weyl-square theory.

We can now expand the Weyl action (\ref{wa}) around the dS vacuum. At quadratic order in $h_{\mu\nu}=g_{\mu\nu}-\bar{g}_{\mu\nu}$, where $\bar{g}$ is the background de Sitter metric and in the transverse-traceless gauge 
\begin{eqnarray}
h^\mu_\mu=\nabla^\mu h_{\mu\nu}=0, 
\end{eqnarray}
we find
\begin{eqnarray}
{\cal S}_{\rm W}&=&-\frac{1}{2g_{W^2}^2}
\int {\rm d}^4 x\Big(\Box h^{\mu\nu}\Box h_{\mu\nu}-6 H^2h^{\mu\nu}
\Box h_{\mu\nu}\nonumber \\
&+&8 H^4 h^{\mu\nu}h_{\mu\nu}\Big). \label{pp}
\end{eqnarray}
The equation obeyed by the spin-2 fluctuations are then
\begin{eqnarray}
\Box^2 h_{\mu\nu}-6 H^2
\Box h_{\mu\nu}+8 H^4 h_{\mu\nu}=0.
\end{eqnarray}
We can write the above equation as 
\begin{eqnarray}
\Big(\Box-4 H^2\Big)\Big(\Box-2H^2\Big)h_{\mu\nu}=0.
\end{eqnarray}
Therefore there are two solutions to the above equation, namely 
\begin{eqnarray}
&&\Big(\Box-2 H^2\Big)h_{\mu\nu}=0, \label{eq1}\\
&&\Big(\Box-4 H^2\Big)h_{\mu\nu}=0. \label{eq2}
\end{eqnarray}
The first equation (\ref{eq1}) describes a massless 
helicity-2 state in 
de Sitter, whereas the second one (\ref{eq2}) describes a partially massless spin-2 state of mass $m^2=2H^2$. 
However, one of these states is a ghost. 

This can also be confirmed by the quadratic action (\ref{pp}). 
Indeed, let us write the action (\ref{pp}) as
\begin{eqnarray}
{\cal S}_{\rm W}
&=&
\frac{1}{2g_{W^2}^2}
\int {\rm d}^4 x\Bigl\{2H^2 
h^{\mu\nu}\Big(\Box-2H^2\Big)h_{\mu\nu}
\nonumber \\
&-&
h^{\mu\nu} \Big(\Box -2 H^2\Big)^2 h_{\mu\nu}\Bigr\}. \label{s1}
\end{eqnarray}
Introducing a Lagrange multiplier tensor field $w_{\mu\nu}$ we may 
write (\ref{s1}) as follows
\begin{eqnarray}
{\cal S}_{\rm W}&=&\frac{1}{2g_{W^2}^2}
\int {\rm d}^4 x\Bigl(2H^2 
h^{\mu\nu}\Big(\Box-2H^2\Big)h_{\mu\nu}+w^{\mu\nu} \Big(\Box \nonumber\\
&-&2 H^2\Big) h_{\mu\nu}+\frac{1}{4}\Big(w^{\mu\nu}w_{\mu\nu}-w^2\Big)\Bigr). \label{s2}
\end{eqnarray}
Defining now new fields 
\begin{eqnarray}
\bar{h}_{\mu\nu}=h_{\mu\nu}-\frac{w_{\mu\nu}}{2H^2}, ~~~~
\bar{w}_{\mu\nu}=\frac{w_{\mu\nu}}{2H^2},
\end{eqnarray}
we may express (\ref{s2}) as
\begin{eqnarray}
{\cal S}_{\rm W}&=&\frac{1}{2g_{W^2}^2}
\int {\rm d}^4 x\left\{2H^2 
\bar h^{\mu\nu}\Big(\Box-2H^2\Big)\bar h_{\mu\nu}+\frac{1}{2}\bar w^{\mu\nu} \Big(\Box  \right.
\nonumber\\ &-&\left.2 H^2\Big) \bar w_{\mu\nu}+H^2\Big(\bar w^{\mu\nu}\bar w_{\mu\nu}-\bar w^2\Big)\right\}. \label{s3}
\end{eqnarray}
Therefore, we see that we have a massless helicity-2 state described by $\bar h_{\mu\nu}$ and a partially massless spin-2 ghost state $\bar w_{\mu\nu}$ of mass, which precisely satisfies the Higuchi bound:
\begin{equation}
M_W^2=2H^2\quad{\rm with}\quad M_W=g_{W^2}M_p\, . \label{mww}
\end{equation}
Once again we see that  the simultaneous limit $M_W,H\rightarrow 0$ is at infinite distance  with respect to the Weyl rescalings of the four-dimensional metric.

\subsection{Infinite distance and the Higuchi bound from the Einstein-$W^2$ theory}

Let us consider now the Einstein-Weyl${}^2$ theory 
\begin{eqnarray}
{\cal S}\!&=&\!\!\int {\rm d}^4 x\left\{\!M_p^2\Big(\!R-2\Lambda\!\Big)\phantom{\frac{X}{Y}}
\nonumber \right.\\
&&\left.~~~~~~~~-\!\frac{1}{2g_{W^2}^2} W_{\mu\nu\rho\sigma}
W^{\mu\nu\rho\sigma}\!\right\}. \label{ein+sw}
\end{eqnarray}
This theory can be considered as a massive Weyl${}^2$ theory
\cite{Ferrara:2018wqd,Ferrara:2018wlb} as  the conformal invariance of the Weyl action is broken both by the Einstein term and the cosmological constant $\Lambda$. In order the theory to admit the dS solution (\ref{R3}), we choose 
$\Lambda=3H^2$. Then the theory (\ref{ein+sw}) has as vacuum   solution the dS spacetime  with 
\begin{eqnarray}
R_{\mu\nu}(\bar{g})= 3H^2 \bar{g}_{\mu\nu}.
\end{eqnarray}
Notice that the Einstein term and the cosmological term removes the degeneracy of the conformal Weyl theory, solutions of which are whole conformal classes, i.e., spacetimes that are conformally related to each other.  
Expanding now (\ref{ein+sw}) at quadratic order in 
$h_{\mu\nu}=g_{\mu\nu}-\bar{g}_{\mu\nu}$ we find that 
\begin{eqnarray}
{\cal S}&=&
\int {\rm d}^4 x\left\{\frac{M_p^2}{2}\Big[ h^{\mu\nu}(\Box-2H^2) h_{\mu\nu}\Big]-\frac{1}{2g_{W^2}^2}\Big(\Box h^{\mu\nu}\Box h_{\mu\nu}\nonumber\right. \\
&& \left.
-6 H^2h^{\mu\nu}
\Box h_{\mu\nu}+8 H^4 h^{\mu\nu}h_{\mu\nu}\Big)\right\}, \label{ein+pp}
\end{eqnarray}
which, equivalently can be written as 
\begin{eqnarray}
{\cal S}&=&
\int {\rm d}^4 x\left\{\frac{M_p^2}{2}\Big[ h^{\mu\nu}(\Box-2H^2) h_{\mu\nu}]
+\frac{1}{2g_{W^2}^2}
\left[2H^2 
\right. \right.\nonumber \\
&&\left.\left. h^{\mu\nu}\Big(\Box-2H^2\Big)h_{\mu\nu}-h^{\mu\nu} \Big(\Box -2 H^2\Big)^2 h_{\mu\nu}\right]\right\}\!. \label{sf}
\end{eqnarray}
We can repeat the steps above, introduce a Lagrange multiplier tensor field $w_{\mu\nu}$ and write (\ref{sf})  as
\begin{eqnarray}
{\cal S}&=&
\int {\rm d}^4 x\left\{\frac{M_p^2}{2}\Big[ h^{\mu\nu}(\Box-2H^2) h_{\mu\nu}\Big]+\frac{1}{2g_{W^2}^2}
\left[2H^2 \right.\right. \nonumber  \\
&&
\left.\left. \hspace{1.5cm}h^{\mu\nu}\Big(\Box-2H^2\Big)h_{\mu\nu}+w^{\mu\nu} \Big(\Box -2 H^2\Big) h_{\mu\nu}\right.\right.\nonumber \\
&&\hspace{1.5cm}\left.\left.+\frac{1}{4}\Big(w^{\mu\nu}w_{\mu\nu}-w^2\Big)\right]\right\}. \label{sf1}
\end{eqnarray}
We may  introduce new fields $\bar{h}_{\mu\nu}$ and $\bar{w}_{\mu\nu}$, defined as  
\begin{eqnarray}
h_{\mu\nu}=\bar{h}_{\mu\nu}-\bar{w}_{\mu\nu}, ~~~~w_{\mu\nu}=2
g_{W^2}^2 \bar{M}_p^2 \bar{w}_{\mu\nu},
\end{eqnarray}
where
\begin{eqnarray}
\bar{M}_p^2=M_p^2+\frac{2H^2}{g_{W^2}^2}, \label{mmw}
\end{eqnarray}
is a new effective Planck mass square. 
Then, the quadratic action (\ref{sf1}) turns out to be
\begin{eqnarray}
{\cal S}&=&
\int {\rm d}^4 x \left\{\frac{\bar{M}_p^2}{2}\Big[ \bar{h}^{\mu\nu}(\Box-2H^2) \bar{h}_{\mu\nu}\Big]-\frac{\bar{M}_p^2}{2}\Big[ \bar{w}^{\mu\nu}(\Box\right. \nonumber \\&&
\left.-2H^2) \bar{w}_{\mu\nu}\Big]+
\frac{g_{W^2}^2\bar{M}_p^4}{2} \Big(\bar{w}^{\mu\nu}\bar{w}_{\mu\nu}-
\bar{w}^2\Big)\right\}. \label{sf2}
\end{eqnarray}
This is the action of a massless helicity-2 state 
$\bar{h}_{\mu\nu}$ and a spin-2 ghost $w_{\mu\nu}$  of mass 
\begin{eqnarray}
M_W=g_{W^2}\bar{M}_p,
\end{eqnarray}
where the effective Planck mass is now $\bar{M}_p^2$ which is $M_p^2$  shifted by $2H^2/g_{W^2}^2$. Therefore, the mass square of the spin-2 state 
$\bar{w}_{\mu\nu}$,
using (\ref{mmw})  is 
\begin{eqnarray}
M_W^2=g_{W^2}^2 M_p^2+2H^2> 2H^2.
\end{eqnarray}
Hence, the Higuchi bound for a spin-s massive state
\begin{eqnarray}
M_W^2\geq s(s-1) H^2,
\end{eqnarray}
 is clearly satisfied. In other words, the spectrum of Einstein-Weyl${}^2$ theory belongs to the complementary series of UIRs of the de Sitter group $SO(1,4)$. Since, the limt $g_{W^2}^2\rightarrow 0$ is  at infinite distance,  
 it follows that saturating the Higuchi bound   \cite{Higuchi:1986py}  is also at infinite distance. Therefore,
 all massive states of the Einstein-Weyl${}^2$ theory that belong to this UIRs, 
 are accumulated at 
 $\sqrt{2}H$.
 Furthermore these states become light
  in the $H\to 0$ limit, in agreement   with the spin-2 swampland conjecture.

%\newline
%\newline

\section{Gravitational Threshold Corrections}

The infinite distance behaviour with respect to Weyl rescaing of the external metric is often related to infinite distance in internal space,  such that extra dimensions are opening up in the infinite distance limit.
This is in fact well for the $U(1)$ gauge coupling constants, where threshold effects from the internal KK modes provide an infinite contribution to the inverse gauge coupling in the decompatification limit.
Concerning the AdS distance conjecture the limit of $\Lambda\rightarrow 0$ is also accompanied by a tower of massless states from an extra compact space \cite{Lust:2019zwm}, as it is true for the KK modes in
the well-known $AdS_d\times S^{d'}$ superstring vacua.

Here we want to investigate in more detail the infinite distance behaviour of the Weyl-square coupling $g_{W^2}$. As will we discuss, one-loop threshold corrections 
to the $W^2$-action provide contributions
to $1/g_{W^2}^2$ which become very large in the decompactification limit of the internal compact space. 
Therefore at one-loop level $g_{W^2}$ goes to zero, when summing over an infinite number of light KK states. This behaviour is consistent with the proposal that coupling constants are emergent, i.e.
go to zero at the boundary of the moduli space, assuming that we can neglect a tree level contribution.

The best-known examples of threshold corrections to the Weyl-square coupling $g_{W^2}$ are provided by ${\cal N}=2$ heterotic or type II string vacua. Here the ${\cal N}=2$ BPS states 
 play an important role in the computation of the one-loop threshold corrections. The ${\cal N}=2$ gauge couplings are given in terms of the ${\cal N}=2$ prepotential.
 Moreover in ${\cal N}=2$ supergravity the $W^2$ action resided in the square of the chiral Weyl superfield. Its  coupling to the abelian vector multiplets is governed by a holomorphic function 
 ${\cal F}_{grav}$,  which at one loop is related to the topological string partition function \cite{Bershadsky:1993ta,Antoniadis:1993ze},
and  whose real part, up to some non-holomorphic corrections is just the Weyl-square coupling $g_{W^2}$, i.e. $g_{W^2}=\Re({\cal F}_{grav})+\dots$.
 
 Typically ${\cal F}_{grav}$ diverges at the boundary of the moduli space due to the sum over an infinite number of light states.
 Consider e.g. the heterotic string on $K_3\times T^2$, where the two-torus is described by the standard moduli fields $T$ and $U$. In addition $S$ denotes the heterotic dilaton.
 Then summing over the ${\cal N}=2$ BPS states, namely KK and winding modes, the holomorphic function ${\cal F}_{grav}$ is given by the following expression  (see e.g.  \cite{LopesCardoso:1995qa}):
 \begin{equation}
{\cal F}_{grav}=S+{b_{grav}\over 8\pi^2}\log \eta^{-2}(T)\eta^{-2}(U)+\dots \, ,
\end{equation}
where $b_{grav}=46+2(n_H-n_V)$. In the large moduli limit, the Dedekind $\eta$ function behaves like $\eta(T)\rightarrow e^{- i\pi T/12}$.
It follows that 
  \begin{equation}
\lim_{T\rightarrow\infty}{\cal F}_{grav}=S-i{b_{grav}\over 48\pi}(T+U)\, .
\end{equation}
So we see that for large $T$ and large $U$, $1/g_{W^2}^2$ diverges, in agreement with the observation that this limit is at large distance in the internal $T^2$ moduli space.

Let us mention at the end of this section that the topological string partition function ${\cal F}_{grav}$ seems to be closely related to the entropy functional ${\cal F}$, which we discussed in section 2.2.2.
Using $S\simeq {1/g_s^2}$ and $T\simeq i R^2$,
${\cal F}_{grav}$ becomes ${\cal F}_{grav}\simeq {1\over g_s^2}+{b_{grav}\over 48\pi}R^2$. Therefore $1/{\cal F}_{grav}\rightarrow 0$ at weak string coupling or for large radius and the distance 
$\Delta\simeq \log {\cal F}_{grav}$ becomes infinite in this limit.

\section{Summary}

In the first part of the paper we were discussing a close relation between the infinite distance conjecture in the context of the swampland hypothesis in quantum gravity and 
the geometric Ricci flow equations. One of the main points of the paper is the conjecture that fixed points of the Ricci flow that appear at infinite distance in the space of backgrounds metrics
have to be accompanied by an infinite tower of massless states. Generically the considered  infinite distance fixed points  belong to Ricci flows
from a non-flat space towards their corresponding flat-space limit. Since the non-flat spaces need an additional  tower of extra states in addition to their zero modes, 
they typically belong to the swampland, as long as they are not augmented by an extra space factor.
Note that the transition to the end point of the flow is somehow discontinuous, since the background at the end point normally has no  infinite tower of states.

As mentioned before in the literature, the Ricci flow is closely related to the RG flow with respect to the energy scale in the underlying two-dimensional non-linear $\sigma$-models.
Hence this connection, when flowing from off-shell string backgrounds towards the on-shell zeroes of the two-dimensional $\beta$-functions, provides an interesting connection between RG and the swampland
idea. This kind of connection was also recently noted in \cite{Gomez:2019ltc}.

The tower of massless states, which appear at the infinite distance fixed points of the Ricci flow, can often be identified with the KK modes or other states of the extra space-factor that is needed in order
to get a consistent string background.
Moreover, 
the Ricci-flow as well as the more generalized gradient flows also provide very useful entropy functionals,  denoted by ${\cal F}$ and ${\cal W}$, which themselves lead to sensible definitions for the related  distances in field space.
These entropy functionals should be  closely related to the degeneracy of the tower of states, which appear in the infinite distance limit.
It is then tempting to speculate that these
entropy functionals  together with their associated tower of light or massless states are in very close relation with the micro states in quantum gravity, in particular  for backgrounds with horizons, like for de Sitter space.

Moreover, it would be very interesting to see if there is  a possible relation between the tower of states in the context of  the infinite distance conjecture   to the soft  graviton modes and their
associated 
asymptotic symmetry groups on the horizon of these backgrounds. For finite horizon, the BMS-like modes should be massive, but in the infinite distance limit of small cosmological constant or of small
curvature, the horizon should approach null infinity and the 
massive modes become massless, just like the massless soft graviton modes. So
in the flat space limit, these modes could possibly become the standard massless soft graviton modes
of the infinite BMS symmetry group at null infinity. 
But to understand how the BMS  modes get a mass, when space is curved, is crucial, and furthermore one has
to understand why the massive soft modes on the finite horizon scale in the same way as the tower of modes in the context of
the large distance conjecture.
This possible complementary  explanation of the tower is analogous to the way when one tries to explain the microscopic degrees of freedom
of a (Schwarzschild) black hole in terms of BMS-like modes - see  e.g. \cite{Hawking:2015qqa,Dvali:2015rea,Hawking:2016msc,Averin:2016ybl,Averin:2016hhm,Hawking:2016sgy,Haco:2018ske}.
For this we refer to some possible future work in progress \cite{workprogress}.

\vskip0.3cm

In the second part of the paper we considered the generalized distance conjecture
for the couplings of the higher curvature interactions. 
We saw that the weak coupling limit is at infinite distance with respect to Weyl rescalings, and hence this limit should be accompanied by an infinite tower of massless states.
  For the quadratic Weyl-square theory this result agrees with the swampland spin-two conjecture and for the $R^2$ we rederive the  (A)dS distance conjecture in this way.

In general the infinite distance conjecture applied to higher curvature theories shows that the limit, when one turns off the higher derivative couplings, belongs to the swampland.
This confirms the general point of view that quantum gravity is only consistent when including an infinite number of higher curvature terms in the effective action.

\vskip0.5cm

\vspace{10px}
{\bf Acknowledgements}
\noindent

We thank  C.~Gomez, E.~Palti, K.~Sfetsos and C.~Vafa     for very useful discussions. 
The work of D.L. is supported  the Origins Excellence Cluster.
The work of S.L.\ is supported by the ERC Starting Grant 679278 Emergent-BH and the ERC Consolidator Grant 772408-Stringlandscape.


\begin{thebibliography}{99}
 
 
 %\cite{Vafa:2005ui}
\bibitem{Vafa:2005ui}
  C.~Vafa,
  ``The String landscape and the swampland,''
  hep-th/0509212.
  %%CITATION = HEP-TH/0509212;%%
  %342 citations counted in INSPIRE as of 23 Aug 2019
 
 %\cite{Ooguri:2006in}
\bibitem{Ooguri:2006in}
  H.~Ooguri and C.~Vafa,
  ``On the Geometry of the String Landscape and the Swampland,''
  Nucl.\ Phys.\ B {\bf 766} (2007) 21
  doi:10.1016/j.nuclphysb.2006.10.033
  [hep-th/0605264].
  %%CITATION = doi:10.1016/j.nuclphysb.2006.10.033;%%
  %283 citations counted in INSPIRE as of 23 Aug 2019
  
    %\cite{Palti:2019pca}
\bibitem{Palti:2019pca}
  E.~Palti,
  ``The Swampland: Introduction and Review,''
  Fortsch.\ Phys.\  {\bf 67} (2019) no.6,  1900037
  doi:10.1002/prop.201900037
  [arXiv:1903.06239 [hep-th]].
  %%CITATION = doi:10.1002/prop.201900037;%%
  %63 citations counted in INSPIRE as of 23 Aug 2019


  
  
  %\cite{ArkaniHamed:2006dz}
\bibitem{ArkaniHamed:2006dz}
  N.~Arkani-Hamed, L.~Motl, A.~Nicolis and C.~Vafa,
  ``The String landscape, black holes and gravity as the weakest force,''
  JHEP {\bf 0706} (2007) 060
  doi:10.1088/1126-6708/2007/06/060
  [hep-th/0601001].
  %%CITATION = doi:10.1088/1126-6708/2007/06/060;%%
  %531 citations counted in INSPIRE as of 23 Aug 2019
  
  
  %\cite{Klaewer:2016kiy}
\bibitem{Klaewer:2016kiy}
  D.~Klaewer and E.~Palti,
  ``Super-Planckian Spatial Field Variations and Quantum Gravity,''
  JHEP {\bf 1701} (2017) 088
  doi:10.1007/JHEP01(2017)088
  [arXiv:1610.00010 [hep-th]].
  %%CITATION = doi:10.1007/JHEP01(2017)088;%%
  %103 citations counted in INSPIRE as of 08 Oct 2019
  
  
%\cite{Heidenreich:2017sim}
\bibitem{Heidenreich:2017sim}
  B.~Heidenreich, M.~Reece and T.~Rudelius,
  ``The Weak Gravity Conjecture and Emergence from an Ultraviolet Cutoff,''
  Eur.\ Phys.\ J.\ C {\bf 78} (2018) no.4,  337
  doi:10.1140/epjc/s10052-018-5811-3
  [arXiv:1712.01868 [hep-th]].
  %%CITATION = doi:10.1140/epjc/s10052-018-5811-3;%%
  %33 citations counted in INSPIRE as of 26 Sep 2019
  
  
  %\cite{Andriolo:2018lvp}
\bibitem{Andriolo:2018lvp}
  S.~Andriolo, D.~Junghans, T.~Noumi and G.~Shiu,
  ``A Tower Weak Gravity Conjecture from Infrared Consistency,''
  Fortsch.\ Phys.\  {\bf 66} (2018) no.5,  1800020
  doi:10.1002/prop.201800020
  [arXiv:1802.04287 [hep-th]].
  %%CITATION = doi:10.1002/prop.201800020;%%
  %48 citations counted in INSPIRE as of 08 Oct 2019
  
  
    %18 citations counted in INSPIRE as of 26 Sep 2019
%\cite{Grimm:2018ohb}
\bibitem{Grimm:2018ohb}
  T.~W.~Grimm, E.~Palti and I.~Valenzuela,
  ``Infinite Distances in Field Space and Massless Towers of States,''
  JHEP {\bf 1808} (2018) 143
  doi:10.1007/JHEP08(2018)143
  [arXiv:1802.08264 [hep-th]].
  %%CITATION = doi:10.1007/JHEP08(2018)143;%%
  %69 citations counted in INSPIRE as of 26 Sep 2019
  
  
  %\cite{Heidenreich:2018kpg}
\bibitem{Heidenreich:2018kpg}
  B.~Heidenreich, M.~Reece and T.~Rudelius,
  ``Emergence of Weak Coupling at Large Distance in Quantum Gravity,''
  Phys.\ Rev.\ Lett.\  {\bf 121} (2018) no.5,  051601
  doi:10.1103/PhysRevLett.121.051601
  [arXiv:1802.08698 [hep-th]].
  %%CITATION = doi:10.1103/PhysRevLett.121.051601;%%
  %56 citations counted in INSPIRE as of 26 Sep 2019
  
  %\cite{Blumenhagen:2018nts}
\bibitem{Blumenhagen:2018nts}
  R.~Blumenhagen, D.~Kl\"awer, L.~Schlechter and F.~Wolf,
  ``The Refined Swampland Distance Conjecture in Calabi-Yau Moduli Spaces,''
  JHEP {\bf 1806} (2018) 052
  doi:10.1007/JHEP06(2018)052
  [arXiv:1803.04989 [hep-th]].
  %%CITATION = doi:10.1007/JHEP06(2018)052;%%
  %43 citations counted in INSPIRE as of 26 Sep 2019

  
  %\cite{Blumenhagen:2018hsh}
\bibitem{Blumenhagen:2018hsh}
  R.~Blumenhagen,
  ``Large Field Inflation/Quintessence and the Refined Swampland Distance Conjecture,''
  PoS CORFU {\bf 2017} (2018) 175
  doi:10.22323/1.318.0175
  [arXiv:1804.10504 [hep-th]].
  %%CITATION = doi:10.22323/1.318.0175;%%
  %34 citations counted in INSPIRE as of 26 Sep 2019


  
  
  %\cite{Lee:2018urn}
\bibitem{Lee:2018urn}
  S.~J.~Lee, W.~Lerche and T.~Weigand,
  ``Tensionless Strings and the Weak Gravity Conjecture,''
  JHEP {\bf 1810} (2018) 164
  doi:10.1007/JHEP10(2018)164
  [arXiv:1808.05958 [hep-th]].
  %%CITATION = doi:10.1007/JHEP10(2018)164;%%
  %46 citations counted in INSPIRE as of 26 Sep 2019
  
  
  %\cite{Lee:2018spm}
\bibitem{Lee:2018spm}
  S.~J.~Lee, W.~Lerche and T.~Weigand,
  ``A Stringy Test of the Scalar Weak Gravity Conjecture,''
  Nucl.\ Phys.\ B {\bf 938} (2019) 321
  doi:10.1016/j.nuclphysb.2018.11.001
  [arXiv:1810.05169 [hep-th]].
  %%CITATION = doi:10.1016/j.nuclphysb.2018.11.001;%%
  %34 citations counted in INSPIRE as of 26 Sep 2019
  
  
  %\cite{Grimm:2018cpv}
\bibitem{Grimm:2018cpv}
  T.~W.~Grimm, C.~Li and E.~Palti,
  ``Infinite Distance Networks in Field Space and Charge Orbits,''
  JHEP {\bf 1903} (2019) 016
  doi:10.1007/JHEP03(2019)016
  [arXiv:1811.02571 [hep-th]].
  %%CITATION = doi:10.1007/JHEP03(2019)016;%%
  %31 citations counted in INSPIRE as of 26 Sep 2019
  
  
  %\cite{Corvilain:2018lgw}
\bibitem{Corvilain:2018lgw}
  P.~Corvilain, T.~W.~Grimm and I.~Valenzuela,
  ``The Swampland Distance Conjecture for K\"ahler moduli,''
  JHEP {\bf 1908} (2019) 075
  doi:10.1007/JHEP08(2019)075
  [arXiv:1812.07548 [hep-th]].
  %%CITATION = doi:10.1007/JHEP08(2019)075;%%

  
    
%\cite{Lee:2019tst}
\bibitem{Lee:2019tst}
  S.~J.~Lee, W.~Lerche and T.~Weigand,
  ``Modular Fluxes, Elliptic Genera, and Weak Gravity Conjectures in Four Dimensions,''
  JHEP {\bf 1908} (2019) 104
  doi:10.1007/JHEP08(2019)104
  [arXiv:1901.08065 [hep-th]].
  %%CITATION = doi:10.1007/JHEP08(2019)104;%%
  %17 citations counted in INSPIRE as of 26 Sep 2019
  
  
  %\cite{Joshi:2019nzi}
\bibitem{Joshi:2019nzi}
  A.~Joshi and A.~Klemm,
  ``Swampland Distance Conjecture for One-Parameter Calabi-Yau Threefolds,''
  JHEP {\bf 1908} (2019) 086
  doi:10.1007/JHEP08(2019)086
  [arXiv:1903.00596 [hep-th]].
  %%CITATION = doi:10.1007/JHEP08(2019)086;%%
  %8 citations counted in INSPIRE as of 26 Sep 2019
  
  
  %\cite{Marchesano:2019ifh}
\bibitem{Marchesano:2019ifh}
  F.~Marchesano and M.~Wiesner,
  ``Instantons and infinite distances,''
  JHEP {\bf 1908} (2019) 088
  doi:10.1007/JHEP08(2019)088
  [arXiv:1904.04848 [hep-th]].
  %%CITATION = doi:10.1007/JHEP08(2019)088;%%
  %8 citations counted in INSPIRE as of 26 Sep 2019


%\cite{Font:2019cxq}
\bibitem{Font:2019cxq}
  A.~Font, A.~Herraez and L.~E.~Ibanez,
  ``The Swampland Distance Conjecture and Towers of Tensionless Branes,''
  JHEP {\bf 1908} (2019) 044
  doi:10.1007/JHEP08(2019)044
  [arXiv:1904.05379 [hep-th]].
  %%CITATION = doi:10.1007/JHEP08(2019)044;%%
  %14 citations counted in INSPIRE as of 26 Sep 2019
  
  %\cite{Lee:2019xtm}
\bibitem{Lee:2019xtm}
  S.~J.~Lee, W.~Lerche and T.~Weigand,
  ``Emergent Strings, Duality and Weak Coupling Limits for Two-Form Fields,''
  arXiv:1904.06344 [hep-th].
  %%CITATION = ARXIV:1904.06344;%%
  %9 citations counted in INSPIRE as of 26 Sep 2019
  
  
 %\cite{Erkinger:2019umg}
\bibitem{Erkinger:2019umg}
  D.~Erkinger and J.~Knapp,
  ``Refined swampland distance conjecture and exotic hybrid Calabi-Yaus,''
  JHEP {\bf 1907} (2019) 029
  doi:10.1007/JHEP07(2019)029
  [arXiv:1905.05225 [hep-th]].
  %%CITATION = doi:10.1007/JHEP07(2019)029;%%
  %3 citations counted in INSPIRE as of 08 Oct 2019

  
  
  
  %\cite{Obied:2018sgi}
\bibitem{Obied:2018sgi}
  G.~Obied, H.~Ooguri, L.~Spodyneiko and C.~Vafa,
  ``De Sitter Space and the Swampland,''
  arXiv:1806.08362 [hep-th].
  %%CITATION = ARXIV:1806.08362;%%
  %254 citations counted in INSPIRE as of 23 Aug 2019
 
  %\cite{Ooguri:2018wrx}
\bibitem{Ooguri:2018wrx}
  H.~Ooguri, E.~Palti, G.~Shiu and C.~Vafa,
  ``Distance and de Sitter Conjectures on the Swampland,''
  Phys.\ Lett.\ B {\bf 788} (2019) 180
  doi:10.1016/j.physletb.2018.11.018
  [arXiv:1810.05506 [hep-th]].
  %%CITATION = doi:10.1016/j.physletb.2018.11.018;%%
  %142 citations counted in INSPIRE as of 22 Aug 2019
  
  %\cite{Garg:2018reu}
\bibitem{Garg:2018reu}
  S.~K.~Garg and C.~Krishnan,
  ``Bounds on Slow Roll and the de Sitter Swampland,''
  arXiv:1807.05193 [hep-th].
  %%CITATION = ARXIV:1807.05193;%%
  %153 citations counted in INSPIRE as of 18 Oct 2019


%\cite{Dvali:2013eja}
\bibitem{Dvali:2013eja}
  G.~Dvali and C.~Gomez,
  ``Quantum Compositeness of Gravity: Black Holes, AdS and Inflation,''
  JCAP {\bf 1401} (2014) 023
  doi:10.1088/1475-7516/2014/01/023
  [arXiv:1312.4795 [hep-th]].
  %%CITATION = doi:10.1088/1475-7516/2014/01/023;%%
  %100 citations counted in INSPIRE as of 26 Sep 2019



%\cite{Dvali:2014gua}
\bibitem{Dvali:2014gua}
  G.~Dvali and C.~Gomez,
  ``Quantum Exclusion of Positive Cosmological Constant?,''
  Annalen Phys.\  {\bf 528} (2016) 68
  doi:10.1002/andp.201500216
  [arXiv:1412.8077 [hep-th]].
  %%CITATION = doi:10.1002/andp.201500216;%%
  %39 citations counted in INSPIRE as of 26 Sep 2019
  
  
  %\cite{Dvali:2017eba}
\bibitem{Dvali:2017eba}
  G.~Dvali, C.~Gomez and S.~Zell,
  ``Quantum Break-Time of de Sitter,''
  JCAP {\bf 1706} (2017) 028
  doi:10.1088/1475-7516/2017/06/028
  [arXiv:1701.08776 [hep-th]].
  %%CITATION = doi:10.1088/1475-7516/2017/06/028;%%
  %35 citations counted in INSPIRE as of 26 Sep 2019

%\cite{Dvali:2018fqu}
\bibitem{Dvali:2018fqu}
  G.~Dvali and C.~Gomez,
  ``On Exclusion of Positive Cosmological Constant,''
  Fortsch.\ Phys.\  {\bf 67} (2019) no.1-2,  1800092
  doi:10.1002/prop.201800092
  [arXiv:1806.10877 [hep-th]].
  %%CITATION = doi:10.1002/prop.201800092;%%
  %58 citations counted in INSPIRE as of 26 Sep 2019


%\cite{Dvali:2018jhn}
\bibitem{Dvali:2018jhn}
  G.~Dvali, C.~Gomez and S.~Zell,
  ``Quantum Breaking Bound on de Sitter and Swampland,''
  Fortsch.\ Phys.\  {\bf 67} (2019) no.1-2,  1800094
  doi:10.1002/prop.201800094
  [arXiv:1810.11002 [hep-th]].
  %%CITATION = doi:10.1002/prop.201800094;%%
  %36 citations counted in INSPIRE as of 26 Sep 2019


%\cite{Lust:2019zwm}
\bibitem{Lust:2019zwm}
  D.~L{\"u}st, E.~Palti and C.~Vafa,
  ``AdS and the Swampland,''
  arXiv:1906.05225 [hep-th].
  %%CITATION = ARXIV:1906.05225;%%
  %1 citations counted in INSPIRE as of 08 Jul 2019
  
  
   %\cite{Klaewer:2018yxi}
\bibitem{Klaewer:2018yxi}
  D.~Klaewer, D.~L{\"u}st and E.~Palti,
  ``A Spin-2 Conjecture on the Swampland,''
  Fortsch.\ Phys.\  {\bf 67} (2019) no.1-2,  1800102
 % doi:10.1002/prop.201800102
  [arXiv:1811.07908 [hep-th]].
  %%CITATION = doi:10.1002/prop.201800102;%%
  %10 citations counted in INSPIRE as of 08 Jul 2019
  

  
  
  %\cite{hamilton}
  \bibitem{hamilton}
  R.S.Hamilton, ``Three manifolds with positive Ricci curvature,'' Jour. Diff. Geom. 17 (1982), 255-306.
  
    %\cite{Perelman:2006un}
\bibitem{Perelman:2006un}
  G.~Perelman,
  ``The Entropy formula for the Ricci flow and its geometric applications,''
  math/0211159 [math-dg].
  %%CITATION = MATH/0211159;%%
  %203 citations counted in INSPIRE as of 22 Aug 2019
  
    \bibitem{sf1}K.~Sfetsos,
  ``Integrable interpolations: From exact CFTs to non-Abelian T-duals,''
  Nucl.\ Phys.\ B {\bf 880}, 225 (2014)
  % doi:10.1016/j.nuclphysb.2014.01.004
  [arXiv:1312.4560 [hep-th]].  
  
\bibitem{sf2}
  K.~Sfetsos and D.~C.~Thompson,
  ``Spacetimes for $\lambda$-deformations,''
  JHEP {\bf 1412} (2014) 164
 % doi:10.1007/JHEP12(2014)164
  [arXiv:1410.1886 [hep-th]].  
  

  
  
  %\cite{Witten:1991yr}
\bibitem{Witten:1991yr}
  E.~Witten,
  ``On string theory and black holes,''
  Phys.\ Rev.\ D {\bf 44} (1991) 314.
  doi:10.1103/PhysRevD.44.314
  %%CITATION = doi:10.1103/PhysRevD.44.314;%%
  %1135 citations counted in INSPIRE as of 30 Sep 2019
  
  
  %\cite{Lust:2019lmq}
\bibitem{Lust:2019lmq}
  D.~L\"ust and E.~Palti,
  ``A Note on String Excitations and the Higuchi Bound,''
  arXiv:1907.04161 [hep-th].
  %%CITATION = ARXIV:1907.04161;%%
  %1 citations counted in INSPIRE as of 30 Sep 2019
  
  
  
  %\cite{Higuchi:1986py}
\bibitem{Higuchi:1986py}
  A.~Higuchi,
  ``Forbidden Mass Range for Spin-2 Field Theory in De Sitter Space-time,''
  Nucl.\ Phys.\ B {\bf 282} (1987) 397.
  doi:10.1016/0550-3213(87)90691-2
  %%CITATION = doi:10.1016/0550-3213(87)90691-2;%%
  %361 citations counted in INSPIRE as of 30 Sep 2019
  

  
  \bibitem{Knopf} 
   B. Chow  and D Knopf, ``The Ricci Flow: An Introduction," AMS, 2004.  
  
  \bibitem {DeWitt}
  B.~S.
  {DeWitt}, {Phys. Rev.} {160}, {1113}  (1967).
  
  
\bibitem{Michor}
{O.}
  {{Gil-Medrano}},  {P.~W.}
  {{Michor}},
  {arXiv:math/9201259 [math.DG]}.
  
  
  %\cite{Bakas:2005kv}
\bibitem{Bakas:2005kv}
  I.~Bakas,
  ``Geometric flows and (some of) their physical applications,''
  Bulg.\ J.\ Phys.\  {\bf 33} (2006) no.s3,  091
  [hep-th/0511057].
  %%CITATION = HEP-TH/0511057;%%
  %9 citations counted in INSPIRE as of 22 Sep 2019
  
  
  %\cite{Bakas:2007qm}
\bibitem{Bakas:2007qm}
  I.~Bakas,
  ``Renormalization group equations and geometric flows,''
  Ann.\ U.\ Craiova Phys.\  {\bf 16} (2006) no.II,  20
  [hep-th/0702034 [HEP-TH]].
  %%CITATION = HEP-TH/0702034;%%
  %10 citations counted in INSPIRE as of 22 Sep 2019





  
  
  
  %\cite{Gomez:2019ltc}
\bibitem{Gomez:2019ltc}
  C.~Gomez,
  ``Gravity as universal UV completion: Towards a unified view of Swampland conjectures,''
  arXiv:1907.13386 [hep-th].
  %%CITATION = ARXIV:1907.13386;%%
  
    %\cite{Larfors:2018nce}
\bibitem{Larfors:2018nce}
  M.~Larfors, A.~Lukas and F.~Ruehle,
  ``Calabi-Yau Manifolds and SU(3) Structure,''
  JHEP {\bf 1901} (2019) 171
  doi:10.1007/JHEP01(2019)171
  [arXiv:1805.08499 [hep-th]].
  %%CITATION = doi:10.1007/JHEP01(2019)171;%%
  %1 citations counted in INSPIRE as of 27 Aug 2019
  
  
  %\cite{Strominger:1986uh}
\bibitem{Strominger:1986uh}
  A.~Strominger,
  ``Superstrings with Torsion,''
  Nucl.\ Phys.\ B {\bf 274} (1986) 253.
  doi:10.1016/0550-3213(86)90286-5
  %%CITATION = doi:10.1016/0550-3213(86)90286-5;%%
  %620 citations counted in INSPIRE as of 27 Aug 2019
  
  %\cite{Hull:1986kz}
\bibitem{Hull:1986kz}
  C.~M.~Hull,
  ``Compactifications of the Heterotic Superstring,''
  Phys.\ Lett.\ B {\bf 178} (1986) 357.
 % doi:10.1016/0370-2693(86)91393-6
  %%CITATION = doi:10.1016/0370-2693(86)91393-6;%%
  %184 citations counted in INSPIRE as of 27 Aug 2019
  
  %\cite{LopesCardoso:2002vpf}
\bibitem{LopesCardoso:2002vpf}
  G.~Lopes Cardoso, G.~Curio, G.~Dall'Agata, D.~L\"ust, P.~Manousselis and G.~Zoupanos,
  ``NonKahler string backgrounds and their five torsion classes,''
  Nucl.\ Phys.\ B {\bf 652} (2003) 5
  doi:10.1016/S0550-3213(03)00049-X
  [hep-th/0211118].
  %%CITATION = doi:10.1016/S0550-3213(03)00049-X;%%
  %304 citations counted in INSPIRE as of 27 Aug 2019

  %\cite{Lust:2004ig}
\bibitem{Lust:2004ig}
  D.~L\"ust and D.~Tsimpis,
  ``Supersymmetric AdS(4) compactifications of IIA supergravity,''
  JHEP {\bf 0502} (2005) 027
  doi:10.1088/1126-6708/2005/02/027
  [hep-th/0412250].
  %%CITATION = doi:10.1088/1126-6708/2005/02/027;%%
  %204 citations counted in INSPIRE as of 29 Aug 2019

  
  
  
  %\cite{Gibbons:1977mu}
\bibitem{Gibbons:1977mu}
  G.~W.~Gibbons and S.~W.~Hawking,
  ``Cosmological Event Horizons, Thermodynamics, and Particle Creation,''
  Phys.\ Rev.\ D {\bf 15} (1977) 2738.
  doi:10.1103/PhysRevD.15.2738
  %%CITATION = doi:10.1103/PhysRevD.15.2738;%%
  %2128 citations counted in INSPIRE as of 22 Aug 2019
  
  
  %\cite{Dvali:2015rea}
\bibitem{Dvali:2015rea}
  G.~Dvali, C.~Gomez and D.~L\"ust,
  ``Classical Limit of Black Hole Quantum N-Portrait and BMS Symmetry,''
  Phys.\ Lett.\ B {\bf 753} (2016) 173
  doi:10.1016/j.physletb.2015.11.073
  [arXiv:1509.02114 [hep-th]].
  %%CITATION = doi:10.1016/j.physletb.2015.11.073;%%
  %37 citations counted in INSPIRE as of 24 Sep 2019
  

 
 
  %\cite{Banks:2000fe}
\bibitem{Banks:2000fe}
  T.~Banks,
  ``Cosmological breaking of supersymmetry?,''
  Int.\ J.\ Mod.\ Phys.\ A {\bf 16} (2001) 910
  doi:10.1142/S0217751X01003998
  [hep-th/0007146].
  %%CITATION = doi:10.1142/S0217751X01003998;%%
  %360 citations counted in INSPIRE as of 22 Aug 2019
  
  
  %\cite{Witten:2001kn}
\bibitem{Witten:2001kn}
  E.~Witten,
  ``Quantum gravity in de Sitter space,''
  hep-th/0106109.
  %%CITATION = HEP-TH/0106109;%%
  %599 citations counted in INSPIRE as of 22 Aug 2019
  
  
   %142 citations counted in INSPIRE as of 22 Aug 2019
  
%\cite{Alvarez-Gaume:2015rwa}
\bibitem{Alvarez-Gaume:2015rwa}
  L.~Alvarez-Gaume, A.~Kehagias, C.~Kounnas, D.~L\"ust and A.~Riotto,
  ``Aspects of Quadratic Gravity,''
  Fortsch.\ Phys.\  {\bf 64} (2016) no.2-3,  176
  doi:10.1002/prop.201500100
  [arXiv:1505.07657 [hep-th]].
  %%CITATION = doi:10.1002/prop.201500100;%%
  %91 citations counted in INSPIRE as of 29 Aug 2019
  
%\cite{Starobinsky:1979ty}
\bibitem{Starobinsky:1979ty}
  A.~A.~Starobinsky,
  ``Spectrum of relict gravitational radiation and the early state of the universe,''
  JETP Lett.\  {\bf 30} (1979) 682
   [Pisma Zh.\ Eksp.\ Teor.\ Fiz.\  {\bf 30} (1979) 719].
  %%CITATION = JTPLA,30,682;%%
  %1416 citations counted in INSPIRE as of 29 Aug 2019  
   
  \bibitem{Bergshoeff:2009hq} 
  E.~A.~Bergshoeff, O.~Hohm and P.~K.~Townsend,
  ``Massive Gravity in Three Dimensions,''
  Phys.\ Rev.\ Lett.\  {\bf 102}, 201301 (2009).
  [arXiv:0901.1766 [hep-th]].
  
  
  %\cite{Gording:2018not}
\bibitem{Gording:2018not}
  B.~Gording and A.~Schmidt-May,
  ``Ghost-free infinite derivative gravity,''
  arXiv:1807.05011 [gr-qc].
  %%CITATION = ARXIV:1807.05011;%%
  

\bibitem{Ferrara:2018wqd} 
  S.~Ferrara, A.~Kehagias and D.~L{\"u}st,
  ``Aspects of Weyl Supergravity,''
  JHEP {\bf 1808}, 197 (2018)
%  doi:10.1007/JHEP08(2018)197
  [arXiv:1806.10016 [hep-th]].

  
  %\cite{Ferrara:2018wlb}
\bibitem{Ferrara:2018wlb}
  S.~Ferrara, A.~Kehagias and D.~L{\"u}st,
  ``Bimetric, Conformal Supergravity and its Superstring Embedding,''
  JHEP {\bf 1905} (2019) 100
%  doi:10.1007/JHEP05(2019)100
  [arXiv:1810.08147 [hep-th]].
  %%CITATION = doi:10.1007/JHEP05(2019)100;%%
  %5 citations counted in INSPIRE as of 08 Jul 2019
  
  
  %\cite{Bershadsky:1993ta}
\bibitem{Bershadsky:1993ta}
  M.~Bershadsky, S.~Cecotti, H.~Ooguri and C.~Vafa,
  ``Holomorphic anomalies in topological field theories,''
  Nucl.\ Phys.\ B {\bf 405} (1993) 279
   [AMS/IP Stud.\ Adv.\ Math.\  {\bf 1} (1996) 655]
  doi:10.1016/0550-3213(93)90548-4
  [hep-th/9302103].
  %%CITATION = doi:10.1016/0550-3213(93)90548-4;%%
  %378 citations counted in INSPIRE as of 30 Aug 2019
  
  %\cite{Antoniadis:1993ze}
\bibitem{Antoniadis:1993ze}
  I.~Antoniadis, E.~Gava, K.~S.~Narain and T.~R.~Taylor,
  ``Topological amplitudes in string theory,''
  Nucl.\ Phys.\ B {\bf 413} (1994) 162
  doi:10.1016/0550-3213(94)90617-3
  [hep-th/9307158].
  %%CITATION = doi:10.1016/0550-3213(94)90617-3;%%
  %301 citations counted in INSPIRE as of 30 Aug 2019
  
  
  %\cite{LopesCardoso:1995qa}
\bibitem{LopesCardoso:1995qa}
  G.~Lopes Cardoso, G.~Curio, D.~L\"ust, T.~Mohaupt and S.~J.~Rey,
  ``BPS spectra and nonperturbative gravitational couplings in N=2, N=4 supersymmetric string theories,''
  Nucl.\ Phys.\ B {\bf 464} (1996) 18
  doi:10.1016/0550-3213(96)00069-7
  [hep-th/9512129].
  %%CITATION = doi:10.1016/0550-3213(96)00069-7;%%
  %53 citations counted in INSPIRE as of 30 Aug 2019
  
  %\cite{Hawking:2015qqa}
\bibitem{Hawking:2015qqa}
  S.~W.~Hawking,
  ``The Information Paradox for Black Holes,''
  arXiv:1509.01147 [hep-th].
  %%CITATION = ARXIV:1509.01147;%%
  %81 citations counted in INSPIRE as of 24 Sep 2019
  
  %\cite{Hawking:2016msc}
\bibitem{Hawking:2016msc}
  S.~W.~Hawking, M.~J.~Perry and A.~Strominger,
  ``Soft Hair on Black Holes,''
  Phys.\ Rev.\ Lett.\  {\bf 116} (2016) no.23,  231301
  doi:10.1103/PhysRevLett.116.231301
  [arXiv:1601.00921 [hep-th]].
  %%CITATION = doi:10.1103/PhysRevLett.116.231301;%%
  %356 citations counted in INSPIRE as of 24 Sep 2019
  
  
  %\cite{Averin:2016ybl}
\bibitem{Averin:2016ybl}
  A.~Averin, G.~Dvali, C.~Gomez and D.~L\"ust,
  ``Gravitational Black Hole Hair from Event Horizon Supertranslations,''
  JHEP {\bf 1606} (2016) 088
  doi:10.1007/JHEP06(2016)088
  [arXiv:1601.03725 [hep-th]].
  %%CITATION = doi:10.1007/JHEP06(2016)088;%%
  %51 citations counted in INSPIRE as of 24 Sep 2019
  
  %\cite{Averin:2016hhm}
\bibitem{Averin:2016hhm}
  A.~Averin, G.~Dvali, C.~Gomez and D.~L\"ust,
  ``Goldstone origin of black hole hair from supertranslations and criticality,''
  Mod.\ Phys.\ Lett.\ A {\bf 31} (2016) no.39,  1630045
  doi:10.1142/S0217732316300457
  [arXiv:1606.06260 [hep-th]].
  %%CITATION = doi:10.1142/S0217732316300457;%%
  %27 citations counted in INSPIRE as of 24 Sep 2019
  

  
  %\cite{Hawking:2016sgy}
\bibitem{Hawking:2016sgy}
  S.~W.~Hawking, M.~J.~Perry and A.~Strominger,
  ``Superrotation Charge and Supertranslation Hair on Black Holes,''
  JHEP {\bf 1705} (2017) 161
  doi:10.1007/JHEP05(2017)161
  [arXiv:1611.09175 [hep-th]].
  %%CITATION = doi:10.1007/JHEP05(2017)161;%%
  %142 citations counted in INSPIRE as of 24 Sep 2019
  
  
  
 
  
    %\cite{Haco:2018ske}
\bibitem{Haco:2018ske}
  S.~Haco, S.~W.~Hawking, M.~J.~Perry and A.~Strominger,
  ``Black Hole Entropy and Soft Hair,''
  JHEP {\bf 1812} (2018) 098
  doi:10.1007/JHEP12(2018)098
  [arXiv:1810.01847 [hep-th]].
  %%CITATION = doi:10.1007/JHEP12(2018)098;%%
  %29 citations counted in INSPIRE as of 24 Sep 2019


 \bibitem{workprogress}
Q.~Bonnefoy, L.~Ciambelli, D.~L\"ust and S.~L\"ust,
``Infinite Black Hole Entropies at Infinite Distances and Tower of States,''
[arXiv:1912.07453 [hep-th]].

 \end{thebibliography}
\end{document}